\documentclass{article}
\usepackage{amssymb,amsfonts,amsmath,graphicx,hyperref}
\begin{document}

\title{Ricci Flow and Nonlinear Reaction--Diffusion Systems\\ in Biology, Chemistry, and Physics}
\author{Vladimir G. Ivancevic\thanks{Defence Science \&
Technology Organisation, Australia
(Vladimir.Ivancevic@dsto.defence.gov.au)} \and Tijana T. Ivancevic\thanks{%
Society for Nonlinear Dynamics in Human Factors \& CITECH Research IP Pty Ltd, Adelaide, Australia (tijana.ivancevic@alumni.adelaide.edu.au)}}\date{}
\maketitle

\begin{abstract}
This paper proposes the Ricci--flow equation from Riemannian geometry as a general
geometric framework for various nonlinear reaction--diffusion systems (and
related dissipative solitons) in mathematical biology. More
precisely, we propose a conjecture that any kind of
reaction--diffusion processes in biology, chemistry and physics
can be modelled by the combined geometric--diffusion system. In
order to demonstrate the validity of this hypothesis, we review a
number of popular nonlinear reaction--diffusion systems and try to show that
they can all be subsumed by the presented geometric framework of the
Ricci flow.\\

\end{abstract}


\section{Introduction}

Parabolic \textit{reaction--diffusion} systems are abundant in
mathematical biology. They are mathematical models that describe
how the concentration of one or more substances distributed in
space changes under the influence of two processes: local chemical
reactions in which the substances are converted into each other,
and diffusion which causes the substances to spread out in space.
More formally, they are expressed as semi--linear parabolic
partial differential equations (PDEs, see e.g., \cite{Purwins}).
The evolution of the state vector $\mathbf{u}(\mathbf{x},t)$
describing the concentration of the different reagents is
determined by anisotropic diffusion as well as local reactions:
\begin{equation}
\partial _{t}\mathbf{{u}=D}\Delta \mathbf{{u}+{R}}(\mathbf{u}),\qquad
(\partial _{t}=\partial/\partial t),  \label{re-dif3D}
\end{equation}%
where each component of the state vector $\mathbf{u}(\mathbf{x},t)$
represents the concentration of one substance, $\Delta$ is the standard
Laplacian operator, $\mathbf{D}$ is a symmetric positive--definite matrix of
diffusion coefficients (which are proportional to the velocity of the
diffusing particles) and $\mathbf{{R}({u})}$ accounts for all local
reactions. The solutions of reaction--diffusion equations display a wide
range of behaviors, including the formation of travelling waves and other
self--organized patterns like \textit{dissipative solitons} (DSs).

On the other hand, the \textit{Ricci flow equation} (or, the parabolic
Einstein equation), introduced by R. Hamilton in 1982 \cite{Ham82}, is the
nonlinear heat--like evolution equation\footnote{%
A hot topic in geometric topology is the Ricci flow, a Riemannian
evolution machinery that recently allowed G. Perelman to prove the
celebrated \textit{Poincar\'{e} Conjecture}, a century--old mathematics
problem (and one of the seven Millennium Prize Problems of the Clay
Mathematics Institute) -- and won him the 2006 Fields Medal (which he
declined in a public controversy) \cite{Mackenzie}. The Poincar\'{e}
Conjecture can roughly be put as a question: Is a closed 3--manifold $M$
topologically a sphere if every closed curve in $M$ can be shrunk
continuously to a point? In other words, Poincar\'{e} conjectured: A
simply-connected compact 3--manifold is diffeomorphic to the 3--sphere $%
S^{3} $ (see e.g., \cite{Yau}).}
\begin{equation}
\partial _{t}g_{ij}=-2R_{ij},  \label{RF}
\end{equation}%
for a time--dependent Riemannian metric $g=g_{ij}(t)$ on a smooth real%
\footnote{%
For the related K\"{a}hler--Ricci flow on complex manifolds, see e.g., \cite%
{GCompl,GaneshADG}.} $n-$manifold $M$ with the Ricci curvature tensor $%
R_{ij} $.\footnote{%
This particular PDE (\ref{RF}) was chosen by Hamilton for much the same
reason that A. Einstein introduced the Ricci tensor into his gravitation
field equation,
\begin{equation*}
R_{ij}-\frac{1}{2}g_{ij}R=8\pi T_{ij},
\end{equation*}%
where $T_{ij}$ is the energy--momentum tensor. Einstein needed a symmetric
2--index tensor which arises naturally from the metric tensor $g_{ij}$ and
its first and second partial derivatives. The Ricci tensor $R_{ij}$ is
essentially the only possibility. In gravitation theory and cosmology, the
Ricci tensor has the volume--decreasing effect (i.e., convergence of
neighboring geodesics, see \cite{HawkPenr}).} This equation roughly says
that we can deform any metric on a 2--surface or $n-$manifold by the
negative of its curvature; after \emph{normalization} (see Figure \ref{Ricci}%
), the final state of such deformation will be a metric with constant
curvature. However, this is not true in general since, in addition to the presence of singularities, the limits could be Ricci solitons (see below). The factor of 2 in (\ref{RF}) is more or less arbitrary, but the
negative sign is essential to insure a kind of global \emph{volume
exponential decay},\footnote{%
This complex geometric process is globally similar to a generic exponential
decay ODE:
\begin{equation*}
\dot{x}=-\lambda f(x),
\end{equation*}%
for a positive function $f(x)$. We can get some insight into its solution
from the simple exponential decay ODE,
\begin{equation*}
\dot{x}=-\lambda x\qquad \text{with the solution}\qquad x(t)=x_{0}\mathrm{e}%
^{-\lambda t},
\end{equation*}%
(where $x=x(t)$ is the observed quantity with its initial value $x_{0}$ and $%
\lambda $ is a positive decay constant), as well as the corresponding $n$th
order rate equation (where $n>1$ is an integer),%
\begin{equation*}
\dot{x}=-\lambda x^{n}\qquad \text{with the solution}\qquad \frac{1}{x^{n-1}}%
=\frac{1}{{x_{0}}^{n-1}}+(n-1)\,\lambda t.
\end{equation*}%
} since the Ricci flow equation (\ref{RF}) is a kind of nonlinear
geometric generalization of the standard linear \emph{heat
equation}\footnote{More precisely, the negative sign is to make the equation parabolic so that there is a theory of existence and uniqueness. Otherwise the equation would be backwards parabolic and not have any theory of existence, uniqueness, etc.}
\begin{equation}
\partial _{t}u=\Delta u.  \label{h1}
\end{equation}%
Like the heat equation (\ref{h1}), the Ricci flow equation (\ref{RF}) is
well behaved in forward time and acts as a kind of smoothing operator (but
is usually impossible to solve in backward time). If some parts of a solid
object are hot and others are cold, then, under the heat equation, heat will
flow from hot to cold, so that the object gradually attains a uniform
temperature. To some extent the Ricci flow behaves similarly, so that the
Ricci curvature `tries' to become more uniform \cite{Milnor}, thus
resembling a monotonic \emph{entropy growth},\footnote{%
Note that two different kinds of entropy functional have been introduced
into the theory of the Ricci flow, both motivated by concepts of entropy in
thermodynamics, statistical mechanics and information theory. One is
Hamilton's entropy, the other is Perelman's entropy. While in Hamilton's
entropy, the scalar curvature $R$ of the metric $g_{ij}$ is viewed as the
leading quantity of the system and plays the role of a probability density,
in Perelman's entropy the leading quantity describing the system is the
metric $g_{ij}$ itself. Hamilton established the monotonicity of his entropy
along the volume-normalized Ricci flow on the 2--sphere $S^{2}$ \cite%
{surface}. Perelman established the monotonicity of his entropy along the
Ricci flow in all dimensions \cite{Perel1}.} $\partial _{t}S\geq 0$, which
is due to the positive definiteness of the metric $g_{ij}\geq 0$, and
naturally implying the \emph{arrow of time} \cite{Penr79,GaneshADG,GCompl}.

In a suitable local coordinate system, the Ricci flow equation (\ref{RF})
has a nonlinear heat--type form, as follows. At any time $t$, we can choose
local harmonic coordinates so that the coordinate functions are locally
defined harmonic functions in the metric $g(t)$. Then the Ricci flow takes
the general form (see e.g., \cite{Anderson})
\begin{equation}
\partial _{t}g_{ij}=\Delta _{M}g_{ij}+Q_{ij}(g,\partial g),  \label{RH}
\end{equation}%
where $\Delta _{M}$ is the \textit{Laplace--Beltrami operator}
(\ref{LB}) and $Q=Q_{ij}(g,\partial g)$ is a lower--order term
quadratic in $g$ and its first order partial derivatives $\partial
g$. From the analysis of nonlinear heat PDEs, one obtains
existence and uniqueness of forward--time solutions to the Ricci
flow on some time interval, starting at any smooth initial metric
$g_{0}$.

The quadratic Ricci flow equation (\ref{RH}) is our geometric
framework for
general bio--reaction--diffusion systems, so that the spatio--temporal PDE (%
\ref{re-dif3D}) corresponds to the quadratic Ricci flow PDE
\begin{equation*}
\begin{array}{ccccc}
\partial _{t}\mathbf{u} & \mathbf{=} & \mathbf{D}\Delta \mathbf{u} & \mathbf{%
+} & \mathbf{R}(\mathbf{u}) \\
\updownarrow  &  & \updownarrow  &  & \updownarrow  \\
\partial _{t}g_{ij} & = & \Delta _{M}g_{ij} & + & Q_{ij}(g,\partial g)%
\end{array}%
\end{equation*}
with:
\begin{itemize}
\item the metric $g=g_{ij}$ on an $n-$manifold $M$ corresponding to the $n-$%
dimensional (or $n-$component, or $n-$phase) concentration $\mathbf{u}(%
\mathbf{x},t)$;

\item the Laplace--Beltrami differential operator $\Delta _{M}$,
as defined on $C^{2}-$functions on an $n-$manifold $M$, with
respect to the Riemannian metric $g_{ij}$, by
\begin{equation}
\Delta _{M}\equiv \frac{1}{\sqrt{\det (g)}}\frac{\partial }{\partial x^{i}}%
\left( \sqrt{\det (g)}g^{ij}\frac{\partial }{\partial
x^{j}}\right) \label{LB}
\end{equation}%
-- $\text{corresponding to the $n-$dimensional bio--diffusion term~~}\mathbf{%
D}\Delta \mathbf{u}$; ~~and

\item the quadratic $n-$dimensional Ricci--term,
$Q=Q_{ij}(g,\partial g)$,
corresponding to the $n-$dimensional bio--reaction term, $\mathbf{R}(\mathbf{%
u})$.
\end{itemize}

As a simple example of the Ricci flow equations (\ref{RF})--(\ref{RH}),
consider a round spherical boundary $S^2$ of the 3--ball radius $r$. The
metric tensor on $S^2$ takes the form
\begin{equation*}
g_{ij}=r^{2}\hat{g}_{ij},
\end{equation*}%
where $\hat{g}_{ij}$ is the metric for a unit sphere, while the Ricci tensor
\begin{equation*}
R_{ij}=(n-1)\hat{g}_{ij}
\end{equation*}%
is independent of $r$. The Ricci flow equation on $S^2$ reduces to
\begin{equation*}
\dot{r}^{2}=-2(n-1),\qquad \text{with the solution\qquad }%
r^{2}(t)=r^{2}(0)-2(n-1)t.
\end{equation*}
Thus the boundary sphere $S^2$ collapses to a point in finite time (see \cite%
{Milnor}).

More generally, the geometrization conjecture \cite{Thurston}
holds for any 3--manifold $M$ (see below). Suppose that we start with a compact initial
3--manifold $M_0$ whose Ricci tensor $R_{ij}$ is everywhere positive
definite. Then, as $M_0$ shrinks to a point under the Ricci flow (\ref{RF}),
it becomes rounder and rounder. If we rescale the metric $g_{ij}$ on $M_0$
so that the volume of $M_0$ remains constant, then $M_0$ converges towards
another compact 3--manifold $M_1$ of constant positive curvature (see \cite%
{Ham82}).

In case of even more general $3-$manifolds (outside the class of positive
Ricci curvature metrics), the situation is much more complicated, as various
singularities may arise. One way in which singularities may arise during the
Ricci flow is that a spherical boundary $S^{2}=\partial M$ of an $3-$%
manifold $M$ may collapse to a point in finite time. Such collapses can be
eliminated by performing a kind of `geometric surgery' on the 3--manifold $M$%
, that is a sophisticated sequence of cutting and pasting without
accumulation of time errors\footnote{%
Hamilton's idea was to perform surgery to cut off the singularities and
continue his flow after the surgery. If the flow develops singularities
again, one repeats the process of performing surgery and continuing the
flow. If one can prove there are only a finite number of surgeries in any
finite time interval, and if the long-time behavior of solutions of the
Ricci flow (\ref{RF}) with surgery is well understood, then one would be
able to recognize the topological structure of the initial manifold. Thus
Hamilton's program, when carried out successfully, would lead to a proof of
the Poincar\'{e} Conjecture and Thurston's Geometrization Conjecture \cite%
{Yau}.} (see \cite{Perel2}). After a finite number of such surgeries, each
component either: (i) converges towards a 3--manifold of constant positive
Ricci curvature which shrinks to a point in finite time, or possibly (ii)
converges towards an $S^{2}\times S^{1}$ which shrinks to a circle $S^{1}$
in finite time, or (iii) admits a `thin--thick' decomposition of \cite%
{Thurston}. Therefore, one can choose the surgery parameters so that there
is a well defined Ricci flow with surgery, that exists for all time \cite%
{Perel2}.

In this paper we use the evolving $n-$dimensional geometric machinery of the
volume--decaying and entropy--growing Ricci flow $g(t)$, given by equations (%
\ref{RF})--(\ref{RH}), for modelling various biological reaction--diffusion
systems and dissipative solitons, defined by special cases of the general
spatio--temporal model (\ref{re-dif3D}).

\section{Bio--reaction--diffusion systems}
\label{r-diff}

In case of ideal mixtures, the driving force for the general diffusion $\mathbf{D}\Delta \mathbf{u}$ (\ref%
{re-dif3D}) is the concentration gradient $-\nabla \bf u$, or the
gradient of the chemical potential $-\nabla u_{i}$ of each species
$u_{i},~(i=1,...,n)$,
giving the \textit{diffusion flux} by the First Fick's law,%
\begin{equation}
J=-\mathbf{D}\nabla \mathbf{u}.  \label{Fick1}
\end{equation}%
Assuming the diffusion coefficients $\mathbf{D}$ to be a constant, the
Second Fick's law gives the linear parabolic heat equation,%
\begin{equation}
\partial _{t}\mathbf{{u}=D}\Delta \mathbf{u},  \label{heat1}
\end{equation}%
while, in case of variable diffusion coefficients $\mathbf{D},$ we get
(slightly) more general parabolic \textit{diffusion equation},%
\begin{equation}
\partial _{t}\mathbf{{u}=\nabla \cdot }\left( \mathbf{D}\nabla \mathbf{u}%
\right),  \label{difeq}
\end{equation}
which is still analogous to the `linear' part of the quadratic
Ricci flow equation (\ref{RH}), $$\partial _{t}g_{ij}=\Delta
_{M}g_{ij},$$ due to general `diffusion properties' of the
Laplace--Beltrami operator $\Delta _{M}$.

The $n-$dimensional diffusion coefficient
$\mathbf{D}=\mathbf{D}(T)$ at different temperatures $T$ can be
approximated by the Arrhenius exponential--decay relation,%
\begin{equation*}
\mathbf{D}(t)=\mathbf{D}_{0}\,\mathrm{e}^{-\frac{E_{A}}{rT}},
\end{equation*}%
where $\mathbf{D}_{0}$ is the maximum possible diffusion
coefficient (at infinite temperature $T$), $E_{A}$ is the
activation energy for diffusion (i.e., the
energy that must be overcome in order for a chemical reaction to occur) and $%
r$ is the gas constant.

Using the First Fick's first law (\ref{Fick1}), the diffusion equation (\ref{difeq}%
)\ can be derived in a straightforward way from the
\emph{continuity equation}, which states that a change in density
in any part of the system is due to inflow and outflow of material
into and out of that part of the system
(effectively, no material is created or destroyed),%
\begin{equation*}
\partial _{t}\textbf{u}+\nabla \cdot \mathbf{j}=0,
\end{equation*}%
where $\mathbf{j}$\ is the flux of the diffusing material.

The most important special case of (\ref{heat1}) is at a steady state, when
the concentrations $\mathbf{u}$ do not change in time, giving the \textit{%
Laplace's equation},
\begin{equation}
\Delta \mathbf{u}=0,\qquad \text{or\qquad }\Delta u_{i}=0,  \label{lap}
\end{equation}%
for harmonic functions $\mathbf{u}=\{u_{i}\}$.

The stochastic version of the deterministic heat equation (\ref{heat1}),
connected with the study of Brownian motion,\footnote{%
Brownian motion is the random movement of particles suspended in a liquid or
gas or the mathematical model used to describe such random movements, often
called a particle theory. The infinitesimal generator (and hence
characteristic operator) of a Brownian motion on $\mathbb{R}^{n}$ is $\frac{1%
}{2}\Delta $, where $\Delta $ is the Laplacian on $\mathbb{R}^{n}.$ More
generally, a Brownian motion on an $n-$manifold $M$ is given by one-half of
the Laplace--Beltrami operator $\Delta_M$ (\ref{LB}).} is the \textit{%
Fokker--Planck equation} (see e.g., \cite{Complexity}),
\begin{equation}
\partial _{t}f=-\partial _{x^{i}}\left[ D_{i}^{1}(x^{i})f\right] +\partial
_{x^{i}x^{j}}\left[ D_{ij}^{2}(x^{i})f\right] ,  \label{Fokker}
\end{equation}%
($\partial _{x^{i}}=\frac{\partial }{\partial x_{i}},$ $\partial
_{x^{i}\,x^{j}}=\frac{\partial ^{2}}{\partial x_{i}\,\partial x_{j}}$),
where where $D_{i}^{1}$ is the drift vector and $D_{ij}^{2}$ the diffusion
tensor (which results from the presence of the stochastic force). The
Fokker--Planck equation (\ref{Fokker}) is used for computing the probability
densities of stochastic differential equations.\footnote{%
Consider the It\={o} stochastic differential equation,%
\begin{equation*}
\mathrm{d}\mathbf{X}_{t}=\boldsymbol{\mu}(\mathbf{X}_{t},t)\,\mathrm{d}t+%
\boldsymbol{\sigma}(\mathbf{X}_{t},t)\,\mathrm{d}\mathbf{W}_{t},
\end{equation*}%
where $\mathbf{X}_{t}\in \mathbb{R}^{n}$ is the state of an $n-$dimensional
stochastic system at time $t$ and $\mathbf{W}_{t}\in \mathbb{R}^{m}$\ is the
standard $m $D Wiener process. If the initial distribution is $\mathbf{X}%
_{0}\sim f(\mathbf{x},0)$, then the probability density of the state is
given by the Fokker--Planck equation (\ref{Fokker}) with the drift and
diffusion terms,
\begin{equation*}
D_{i}^{1}(\mathbf{x},t)=\mu _{i}(\mathbf{x},t)\qquad \text{and}\qquad
D_{ij}^{2}(\mathbf{x},t)=\frac{1}{2}\sum_{k}\sigma _{ik}(\mathbf{x},t)\sigma
_{kj}^{\mathsf{T}}(\mathbf{x},t).
\end{equation*}%
}

Also, notice that the real--valued heat equation (\ref{heat1}) is formally
similar to the complex--valued \textit{Schr\"{o}dinger equation} (see e.g., \cite%
{QuLeap}),
\begin{equation}
\partial _{t}\psi =\frac{i\hbar }{2m}\Delta \psi ,  \label{Schr}
\end{equation}%
where $\psi =\psi (\mathbf{x},t)$ is the wave--function of the particle, $i=%
\sqrt{-1},$ and $\hbar $ is Planck's constant divided by $2\pi .$

In the remainder of this section, we will review a number of
particular bio--reaction--diffusion systems, which are likely to be
subsumed by the quadratic Ricci flow model (\ref{RH}).

\subsection{1--component systems}

\subsubsection{Kolmogorov--Petrovsky--Piscounov equation}

The simplest bio--re-action--diffusion PDE concerning the concentration $%
u=u(x,t)$ of a single substance in one spatial dimension,
\begin{equation}
\partial _{t}u=D\partial _{x}^{2}u+R(u),  \label{re-dif1D}
\end{equation}%
is also referred to as the Kolmogorov--Petrovsky--Piscounov (KPP) equation.
If the reaction term vanishes, then the equation represents a pure diffusion
process described by the heat equation. In particular, the choice
\begin{equation*}
R(u)=u(1-u)
\end{equation*}%
yields \textit{Fisher's equation} that was originally used to describe the
spreading of biological populations.\footnote{%
In addition, the effects of convection and quenched spatial disorder on the
evolution of a population density are described by a generalization of the
Fisher/KPP equation given by \cite{Nelson98}
\begin{equation*}
\partial _{t}u=D\nabla ^{2}u+Uu-qu^{2},
\end{equation*}%
where $u=(\mathbf{x},t)$ represents the population density, $D$ is a
spatially homogenous diffusion constant, $U=U(\mathbf{x})$ is a spatially
inhomogeneous growth term, and $q=b{\ell _{0}}^{d}$ is a competition term ($%
b $ is a competition rate and $\ell _{0}$ is the microscopic length scale at
which two particles will compete with one another) . One simple form of
inhomogeneity considered in these works is a `square well' potential $U(%
\mathbf{x})$ which consists of a uniform space with negative
growth rate (termed the `desert'), in which a single region of
positive growth rate (an `oasis') is placed. This model has proven
to be applicable to experiments with bacteria populations in
adverse environments \cite{Missel}.}

The one--component KPP equation (\ref{re-dif1D}) can also be written in the
variational (gradient) form
\begin{equation}
\partial _{t}u=-\frac{\delta F}{\delta u},  \label{var}
\end{equation}%
and therefore describes a permanent decrease (a kind of exponential decay)
of the system's \emph{free energy} functional
\begin{equation*}
F=\int\limits_{-\infty }^{\infty }\left[ \frac{D}{2}(\partial _{x}u)^{2}+V(u)%
\right] dx,
\end{equation*}%
where $V(u)$ is the potential such that
\begin{equation}
R(u)=-\frac{dV(u)}{du}.  \label{ru}
\end{equation}

\subsubsection{Swift--Hohenberg equation}

The Swift--Hohenberg (SH) equation, no-ted for its
pattern--forming behavior, is the decaying reaction--diffusion
PDE,
\begin{equation}
\partial _{t}u=-(1+\Delta )^2u+R(u),  \label{SH}
\end{equation}
given by the variational (gradient) equation (\ref{var}) with the
free energy functional
\begin{equation*}
F=\int\limits_{\Omega }\left[ V(u)+\frac{1}{2}\left( (1+\Delta )u\right) ^{2}%
\right] dxdy,
\end{equation*}
where $R(u)$ is given by (\ref{ru}), while $\Omega $ is a
2--dimensional region in which (bio)chemical pattern formation
occurs.

The time derivative of the free energy $F$ is given by
\begin{equation*}
\partial _{t}F=\int\limits_{\Omega }\left[ \frac{dV(u)}{du}+(1+\Delta )u%
\right] \partial _{t}u\,dxdy,
\end{equation*}
and, since the expression in square brackets is equal to the negative
right--hand side of (\ref{SH}), we have
\begin{equation*}
\dot{F}=-\int\limits_{\Omega }\left( \partial _{t}u\right) ^{2}dxdy\leq 0.
\end{equation*}
Therefore, the free energy $F$ is the \emph{Lyapunov functional} that may
only decrease as it evolves along its trajectory in some phase space. If $F$
has no minima, then when the horizontal scale of the liquid container is
large compared to the instability wavelength, a propagating front will be
observed (e.g., in chemically reacting flames). In this case, $F$ will
decrease continuously until the front approaches the boundary of the medium.
An alternative possibility is realized when $F$ has one or several minima,
each corresponding to a local equilibrium state in time. In this case the
so--called multi--stability is possible. Therefore, the limit behavior of
gradient systems of the form of (\ref{ru}) is characterized by either a
steady attractor or propagating fronts \cite{Rabi}.

\subsubsection{Ginzburg--Landau equation}

One of the most popular models in the pattern--formation theory is
the complex Ginzburg--Landau equation (see e.g., \cite{Rabi}),
\begin{equation}
\partial _{t}A=\varepsilon A+(1+i\alpha )\Delta A-(1+i\beta )|A|^{2}A,
\label{CGL}
\end{equation}
where $A$\ is the complex wave amplitude, $i=\sqrt{-1},
\varepsilon $ is the super-criticality parameter, while $\alpha $
and $\beta $ measure linear and nonlinear dispersion (the
dependence of the frequency of the waves on the wave-number),
respectively. The equation (\ref{CGL}) describes a vast array of
phenomena including nonlinear waves, second-order phase
transitions, Rayleigh--B\'{e}nard convection and
superconductivity. The equation describes the evolution of
amplitudes of unstable modes for any process exhibiting a Hopf
bifurcation, for which a continuous spectrum of unstable
wave--numbers is taken into account. It can be viewed as a highly
general normal form for a large class of bifurcations and
nonlinear wave phenomena in spatially extended
systems.\footnote{The extension of the complex Ginzburg--Landau
equation (\ref{CGL}), which describes strongly resonant
multi--frequency forcing of the form
$$F=f_{1}e^{i\omega\tilde{t}}+f_{2}e^{2i\omega\tilde{t}}+f_{3}e^{3i\omega\tilde{t}}+c.c.$$
was recently proposed in \cite{Conway08} by considering the
analogous center--manifold reduction of the extended dynamical
system in which the forcing amplitudes $f_{1}$, $f_{2}$, and
$f_{3}$ are considered as dynamical variables that vary on the
slow time scale $t$. Under time translations $T_{\tau}:\,
A\rightarrow Ae^{i\omega\tau}$, they transform as
$f_{1}\rightarrow f_{1}e^{i\omega\tau},\, f_{2}\rightarrow
f_{2}e^{2i\omega\tau},\, f_{3}\rightarrow f_{3}e^{3i\omega\tau}$.
To cubic order in $A$ the most general equation that is
equivariant under $T_{\tau}$ is then given by
\begin{equation} \partial_t A=a_{1}+a_{2}A+a_{3}\Delta
A+a_{4}A|A|^{2}+a_{5}\bar{A}+a_{6}\bar{A}^{2},\label{eq:cgle_early}
\end{equation}
 where $a_{1}=b_{11}f_{1}+b_{12}\bar{f}_{2}f_{3}$, $a_{2}=b_{21}+b_{22}|f_{3}|^{2}$,
$a_{5}=b_{51}f_{2}$, $a_{6}=b_{61}f_{3}$. The forcing terms
$f_{j}$ satisfy decoupled evolution equations on their own. In the
simplest case this evolution expresses a de-tuning $\nu_{j}$ of
the forcing $f_{j}$ from the respective resonance and the $f_{j}$
satisfy
\begin{equation*} \dot{f}_{j}=i\nu_{j}f_{j},\qquad
(j=1\ldots 3).
\end{equation*}
In general, the de-tuning introduces time dependence into
(\ref{eq:cgle_early}).}

In particular, if we put $\alpha =\beta =0\ $in (\ref{CGL}), we get the
real, or dissipative, Ginzburg--Landau equation,
\begin{equation}
\partial _{t}A=\varepsilon A+\Delta A-|A|^{2}A,  \label{DGL}
\end{equation}%
which is a gradient equation: $\partial _{t}A=-\delta F/\delta A,$ with the
free energy functional
\begin{equation*}
F=-\int_{\Omega }\left[ \varepsilon |A|^{2}-\frac{1}{2}|A|^{4}+\left( \nabla
A\right) ^{2}\right] dxdy.
\end{equation*}%
Using the fact that
\begin{equation*}
\dot{F}=-\int_{\Omega }|\partial _{t}A|^{2}dxdy\leq 0,
\end{equation*}%
solutions of (\ref{DGL}) at $t\rightarrow \infty $ are either stationary
field--distributions satisfying, for $\varepsilon =1,$ the equation
\begin{equation}
\Delta A+A-|A|^{2}A=0,  \label{DA}
\end{equation}%
of fronts whose propagation is accompanied by a decrease of the functional $%
F $. The functional must reach its minimum at stable stationary solutions of
(\ref{DA}).

\subsubsection{Neural field theory}

The dynamical system from which the temporal evolution of neural
activation fields is generated is constrained by the postulate
that localized peaks of activation are stable objects, or
formally, \textit{fixed--point attractors}. Such a field dynamics
has the generic form \cite{Schoner}
\begin{equation}
\tau \partial_t u = -u + \text{resting level}+\text{input}+\text{interaction,%
}  \label{dft1}
\end{equation}%
where $u=u(x,t)$ is the activation field defined over the metric dimension $%
x $ and time $t.$ The first three terms define an input driven regime, in
which attractor solutions have the form
\begin{equation*}
u(x,t)=\text{resting \thinspace level}+\text{input.}
\end{equation*}
The \textit{rate of relaxation} is determined by the time scale parameter $%
\tau $. The interaction stabilizes localized peaks of activation against
decay by local excitatory interaction and against diffusion by global
inhibitory interaction. In Amari's formulation \cite{Amari} the conceptual
model (\ref{dft1})\ is specified as a \textit{continuous model for neural
activity in cortical structures},
\begin{equation}
\tau \partial_t u(x,t)=-u(x,t)+h+S(x,t)+\int dx^{\prime }w(x-x^{\prime
})\sigma (u(x^{\prime },t)),  \label{am}
\end{equation}%
where $h<0$ is a constant resting level, $S(x,t)$ is spatially and
temporally variable input function, $w(x)$ is an interaction kernel and $%
\sigma (u)$ is a sigmoidal nonlinear threshold function. The
interaction term collects input from all those field sites
$x^{\prime }$ at which activation is sufficiently large. The
interaction kernel determines if inputs from those sites are
positive, driving up activation (excitatory), or negative, driving
down activation (inhibitory). Excitatory input from nearby
location and inhibitory input from all field locations generically
stabilizes localized peaks of activation. For this class of
dynamics, detailed analytical results provide a framework for the
inverse dynamics task facing the modeler, determining a dynamical
system that has the appropriate attractor solutions
\cite{StrAttr,TacaNODY}.\footnote{Recently, a \textit{neural
attractor dynamics} (NAD) was designed (see \cite{Schoner}) based
on a discretization for single neurons of Amari's neural field
equation (\ref{am}). The so--called \textit{discrete Amari
equation} describes the temporal evolution of the activity of all
single neurons considering positive and negative contributions
from external input and internal neural interactions. Since only
activated neurons can have an impact on other neurons, the neural
attractor dynamics is nonlinear, and effects of bi--stability and
hysteresis can be used for low--level memory and neural
competition. The NAD describes the temporal rate of change of the
dynamical variable $u_{i}$ of neural activity for all behavioral
neurons $i.$ It is formulated as the following differential
equation:
\begin{equation}
\tau \dot{u}_{i}=-u_{i}+h+s_{i}^{\text{beh}}+c_{\text{mot}}\cdot
\sigma
(m_{i})+\alpha _{\text{selfexc},i}^{\text{beh}}+\alpha _{\text{exc},i}^{%
\text{beh}}-\alpha _{\text{inh},i}^{\text{beh}},  \label{disAmEq}
\end{equation}%
where the system parameters have the following meaning:

$\tau ,$ the constant relaxation rate, i.e., the time scale on
which the dynamics reacts to changes;

$h,$ the constant negative resting level of neural activation;

$\sigma (.),$ a sigmoidal function, which maps the value of neural
activity onto $[0,1],$ given by\\ $ \sigma
(u)=\frac{1}{1+\text{e}^{-\beta u}}, $ where $\beta $ (=100)
parameterizes the slope of the resulting function;

$s_{i}^{\text{beh}},$ the adequate stimulus provided by sensory
input of a certain duration;

$u_{i},$ activity of behavioral neuron $i$, i.e., activity of
behavior $i;$

$c_{\text{mot}},$ a constant for weighting the motivational contribution, $%
c_{\text{mot}}<|h|;$

$\alpha _{\text{selfexc},i}^{\text{beh}}$ excitatory contribution of neuron $%
i$'s own activity $u_{i}$;

$\alpha _{\text{exc},i}^{\text{beh}}$, all excitatory contribution
of active neurons connected to neuron $i;$

$\alpha _{\text{inh},i}^{\text{beh}}$, all inhibitory contribution
of active neurons connected to neuron $i$

$m_{i},$ activity of motivational neuron $i,$ i.e., motivation of
behavior $ i$ is in \cite{Schoner} defined by the following
NAD--equation, similar to (\ref{disAmEq}):
\begin{equation*}
\tau \dot{m}_{i}=-m_{i}+h+s_{i}^{\text{mot}}+\alpha _{\text{selfexc},i}^{%
\text{mot}}+\alpha _{\text{exc},i}^{\text{mot}}-\alpha _{\text{inh},i}^{%
\text{mot}},
\end{equation*}%
where

$\alpha _{\text{selfexc},i}^{\text{mot}},$ excitatory contribution of neuron $%
i$'s own motivation $m_{i}$; \qquad

$\alpha _{\text{exc},i}^{\text{mot}}$, all excitatory contribution
of motivation neurons connected to neuron $i;$

$\alpha _{\text{inh},i}^{\text{mot}}$, all inhibitory contribution
of motivation neurons connected to neuron $i.$

In this framework, a nonlinear neural dynamical and control system
generates the temporal evolution of behavioral variables, such
that desired behaviors are fixed-point attractor solutions while
un-desired behaviors are repellers.

This kind of \emph{attractor \& repeller dynamics} \cite{StrAttr}
provides the basis for understanding \emph{cognition}, both
natural and artificial \cite{TacaNODY,NeuFuz,CompMind}.}

\subsection{2--component systems}

Two--component systems allow for a much larger range of possible phenomena
than their one--component counterparts. An important idea that was first
proposed by A. Turing is that a state that is stable in the local system
should become unstable in the presence of diffusion \cite{Turing}. This idea
seems counter--intuitive at first glance as diffusion is commonly associated
with a stabilizing effect. However, the linear stability analysis shows that
when linearizing the general two--component system%
\begin{equation*}
\left(
\begin{array}{c}
\partial _{t}u \\
\partial _{t}v%
\end{array}%
\right) =\left(
\begin{array}{cc}
D_{u} & 0 \\
0 & D_{v}%
\end{array}%
\right) \left(
\begin{array}{c}
\partial _{xx}u \\
\partial _{xx}v%
\end{array}%
\right) +\left(
\begin{array}{c}
F(u,v) \\
G(u,v)%
\end{array}%
\right)
\end{equation*}%
and perturbing the system against plane waves%
\begin{equation*}
\tilde{\mathbf{u}}_{\mathbf{k}}(\mathbf{x},t)=\left(
\begin{array}{c}
\tilde{u}(t) \\
\tilde{v}(t)%
\end{array}
\right) \mathrm{e}^{i\mathbf{k}\cdot \mathbf{x}}
\end{equation*}%
close to a stationary homogeneous solution one finds \cite{CompMind}
\begin{equation*}
\left(
\begin{array}{c}
\partial _{t}\tilde{u}_{\mathbf{k}}(t) \\
\partial _{t}\tilde{v}_{\mathbf{k}}(t)%
\end{array}%
\right) =-k^{2}\left(
\begin{array}{c}
D_{u}\tilde{u}_{\mathbf{k}}(t) \\
D_{v}\tilde{v}_{\mathbf{k}}(t)%
\end{array}%
\right) +\mathbf{R}^\prime \left(
\begin{array}{c}
\tilde{u}_{\mathbf{k}}(t) \\
\tilde{v}_{\mathbf{k}}(t)%
\end{array}%
\right) .
\end{equation*}%
Turing's idea can only be realized in four equivalence classes of systems
characterized by the signs of the Jacobian $\mathbf{R}^\prime $ of the
reaction function. In particular, if a finite wave vector k is supposed to
be the most unstable one, the Jacobian must have the signs%
\begin{equation*}
\left(
\begin{array}{cc}
+ & - \\
+ & -%
\end{array}%
\right) ,\quad \left(
\begin{array}{cc}
+ & + \\
- & -%
\end{array}%
\right) ,\quad \left(
\begin{array}{cc}
- & + \\
- & +%
\end{array}%
\right) ,\quad \left(
\begin{array}{cc}
- & - \\
+ & +%
\end{array}%
\right) .
\end{equation*}%
This class of systems is named activator$-$inhibitor system after its first
representative: close to the ground state, one component stimulates the
production of both components while the other one inhibits their growth. Its
most prominent representative is the FitzHugh--Nagumo equation (\ref{FN}).

\subsubsection{Brusselator}

Classical model of an autocatalytic chemical reaction is Prigogine's
Brusselator (see e.g., \cite{Pri80})
\begin{equation}
\partial _{t}u=D_{u}^{2}\Delta u+\alpha +u^{2}v-(1+\beta )u,\qquad \partial
_{t}v=D_{v}^{2}\Delta v-u^{2}v+\beta u,  \label{Brus}
\end{equation}%
which describe the spatio--temporal evolution of the intermediate components
$u$ and $v,$ with diffusion coefficients $D_{u}$ and $D_{v}$, while
reactions
\begin{equation*}
\alpha \overset{r_{1}}{\longrightarrow }u,\qquad 2u+v\overset{r_{2}}{%
\longrightarrow }3u,\qquad \beta +u\overset{r_{3}}{\longrightarrow }%
v+d,\qquad u\overset{r_{4}}{\longrightarrow }c
\end{equation*}%
describe the concentration of the original substances $\alpha $ and $\beta ,$
for which the final products $c$ and $d$ are constant when all four reaction
rates $r_{i}$ equal unity.

A discretized (temporal only) version of the Brusselator PDE (\ref%
{Brus}) reads
\begin{equation*}
\dot{u}=\alpha +u^{2}v-(1+\beta )u,\qquad \dot{v}=\beta u-u^{2}v.
\end{equation*}

The Brusselator displays oscillatory behavior in the species $u$ and $v$
when reverse reactions are neglected and the concentrations of $\alpha$ and $%
\beta$ are kept constant.

\subsubsection{2--component model of excitable media}

Turbulence of scroll waves is a kind of spatio--temporal chaos that exists
in 3--dimensional excitable media. Cardiac tissue and the
Belousov--Zhabotinsky reaction are examples of such media. In cardiac
tissue, chaotic behavior is believed to underlie fibrillation which, without
intervention, precedes cardiac death. Fast computer--simulation of waves in
excitable media have been often performed using the 2--component Barkley
model of excitable media \cite{Barkley},
\begin{equation}
\partial _{t}u=\frac{1}{\epsilon }u(1-u)\left( u-\frac{v+b(t)}{a}\right)
+\nabla ^{2}u+h(t),\qquad \partial _{t}v=u-v,  \label{Bark}
\end{equation}%
where $\epsilon $ is a small parameter $\epsilon \ll 1$ characterising
mutual time scales of the fast $u$ and slow $v$ variables, and $a$ and $b$
specify the kinetic properties of the system. Parameter $b$ determines the
excitation threshold and thus controls the excitability of the medium. The
term $h(t)$ represents an `extra transmembrane current'.

Suppression of the turbulence using stimulation of two different types,
`modulation of excitability' and `extra transmembrane current' was performed
in \cite{Morgan}, using the Barkley model (\ref{Bark}). With cardiac
defibrillation in mind, the authors used a single pulse as well as
repetitive extra current with both constant and feedback controlled
frequency. They show that turbulence can be terminated using either a
resonant modulation of excitability or a resonant extra current. The
turbulence is terminated with much higher probability using a resonant
frequency perturbation than a non-resonant one. Suppression of the
turbulence using a resonant frequency is up to fifty times faster than using
a non-resonant frequency, in both the modulation of excitability and the
extra current modes. They also demonstrate that resonant perturbation
requires strength one order of magnitude lower than that of a single pulse,
which is currently used in clinical practice to terminate cardiac
fibrillation.

\subsubsection{Gierer--Meinhardt activator--inhibitor system}

Spontaneous pattern formation in initially almost homogeneous systems is
common in both organic and inorganic systems. The Gierer--Meinhardt model
\cite{GM72} is a reaction--diffusion system of the activator--inhibitor type
that appears to account for many important types of pattern formation and
morphogenesis observed in biology, chemistry and physics. The model
describes the concentration of a short--range autocatalytic substance, the
activator, that regulates the production of its long--range antagonist, the
inhibitor. It is given as a 2--component nonlinear PDE system,
\begin{equation*}
\partial _{t}a=-\mu _{a}a+\rho a^{2}/h+D_{a}\partial _{x^{2}}a+\rho
_{a},\qquad \partial _{t}h=-\mu _{h}h+\rho a^{2}+D_{h}\partial
_{x^{2}}h+\rho _{h},
\end{equation*}%
where $a$ is a short--range autocatalytic substance, i.e., \textit{activator}%
, and $h$ is its long--range antagonist, i.e., \textit{inhibitor}. $\partial
_{t}a$ and $\partial _{t}h$ describe respectively the changes of activator
and inhibitor concentrations per second, $\mu _{a}$ and $\mu _{h}$\ are the
corresponding decay rates, while $D_{a}$ and $D_{h}$ are the corresponding
diffusion coefficients. $\rho $ is a positive constant. $\rho _{a}$ is a
small activator--independent production rate of the activator and is
required to initiate the activator autocatalysis at very low activator
concentration, e.g., in the case of regeneration. A low baseline production
of the inhibitor, $\rho _{h}$, leads to a stable non--patterned steady
state; the system can be asleep until an external trigger occurs by an
elevation of the activator concentration above a threshold \cite{GM06}.

\subsubsection{Fitzhugh--Nagumo activator--inhibitor system}

An important example of bio--reaction--diffusion systems,
frequently used in neurodynamics, is the
2--component Fitzhugh--Nagumo activator--inhibitor system \cite%
{Fitzhugh,Nagumo} (see also \cite{GaneshWSci,NeuFuz})
\begin{equation}
\tau _{u}\partial _{t}u=D_{u}^{2}\Delta u+f(u)-\sigma v,\qquad \tau
_{v}\partial _{t}v=D_{v}^{2}\Delta v+u-v,  \label{FN}
\end{equation}%
with $f(u)=\lambda u-u^{3}-\kappa $, which describes how an action potential
travels through a nerve, $D_{u}$ and $D_{v}$ are diffusion coefficients, $%
\tau _{u}\ $and $\tau _{v}$ are time characteristics, while $\kappa ,\sigma $
and $\lambda $ are positive constants. In matrix form, system (\ref{FN})
reads
\begin{equation*}
\left(
\begin{array}{c}
\tau _{u}\partial _{t}u \\
\tau _{v}\partial _{t}v%
\end{array}%
\right) =\left(
\begin{array}{cc}
D_{u}^{2} & 0 \\
0 & D_{v}^{2}%
\end{array}%
\right) \left(
\begin{array}{c}
\Delta u \\
\Delta v%
\end{array}%
\right) +\left(
\begin{array}{c}
\lambda u-u^{3}-\kappa -\sigma v \\
u-v%
\end{array}%
\right) .
\end{equation*}%
When an activator--inhibitor system undergoes a change of parameters, one
may pass from conditions under which a homogeneous ground state is stable to
conditions under which it is linearly unstable. The corresponding
bifurcation may be either a \textit{Hopf bifurcation} to a globally
oscillating homogeneous state with a dominant wave number $k=0$ or a \textit{%
Turing bifurcation} to a globally patterned state with a dominant finite
wave number. The latter in two spatial dimensions typically leads to stripe
or hexagonal patterns.

In particular, for the Fitzhugh--Nagumo system (\ref{FN}), the neutral
stability curves marking the boundary of the linearly stable region for the
Turing and Hopf bifurcation are given by
\begin{equation*}
\begin{array}{rrl}
q_{\text{n}}^{H}(k): & \frac{1}{\tau }+(d_{u}^{2}+\frac{1}{\tau }%
d_{v}^{2})k^{2} & =f^{\prime }(u_{h}), \\
q_{\text{n}}^{T}(k): & \frac{\kappa _{3}}{1+d_{v}^{2}k^{2}}+d_{u}^{2}k^{2} &
=f^{\prime }(u_{h}).%
\end{array}%
\end{equation*}%
If the bifurcation is subcritical, often localized structures (i.e.,
dissipative solitons) can be observed in the hysteretic region where the
pattern coexists with the ground state. Other frequently encountered
structures comprise pulse trains, spiral waves and target patterns.

The reduced (temporal) non--dimensional Fitzhugh--Nagumo equations read:
\begin{eqnarray}
\dot{v}\, &=&\,v(a-v)(v-1)-w+I_{a},  \label{fhn1} \\
\dot{w}\, &=&\,bv-\gamma w,  \label{fhn2}
\end{eqnarray}%
where $0<a<1$ is essentially the threshold value, $b$ and $\gamma $ are
positive constants and $I_{a}$ is the applied current. The drift field for
this model is given by
\begin{equation*}
u_{1}(v,w)\,=\,v(a-v)(v-1)-w,\qquad u_{2}(v,w)\,=\,bv-\gamma w.
\end{equation*}

As can be seen from (\ref{fhn2}) the null cline of the
deterministic dynamics of this equations is the line
$v=\frac{\gamma }{b}w$. By substitution on the r.h.s of equation
(\ref{fhn1}) we find the following equation for steady states:~
$
v(a-v)(v-1)-\frac{b}{\gamma }v=0.
$

When this system is in a noisy environment, in the limit of weak noise, we
can approximate the dynamics of the fluctuations by the \emph{Langevin
equation} \cite{GaneshWSci,Complexity}
\begin{equation*}
\dot{v}\,=\,v(a-v)(v-1)-\frac{b}{\gamma }v+\xi (t),
\end{equation*}
that is, the fluctuations run along the line $v=\frac{\gamma }{b}w$.

In particular, parameters in the \textit{FitzHugh--Nagumo neuron} model \cite%
{TacaNODY,CompMind}
\begin{equation*}
\dot{v}=a+bv+cv^{2}+dv^{3}-u,\qquad \dot{u}=\varepsilon (ev-u),
\end{equation*}%
can be tuned so that the model describes spiking dynamics of many resonator
neurons. Since one needs to simulate the shape of each spike, the time step
in the model must be relatively small, e.g., $\tau =0.25\,ms$. Since the
model is a 2--dimensional system of ODEs, without a reset, it cannot exhibit
autonomous chaotic dynamics or bursting. Adding noise to this, or some other
2--dimensional models, allows for stochastic bursting.

\subsubsection{2--component Belousov--Zhabotinsky reaction}

Classical Belousov--Zhabotinsky (BZ) reaction is a family of oscillating
chemical reactions. During these reactions, transition--metal ions catalyze
oxidation of various, usually organic, reductants by bromic acid in acidic
water solution. Most BZ reactions are homogeneous. The BZ reaction makes it
possible to observe development of complex patterns in time and space by
naked eye on a very convenient human time scale of dozens of seconds and
space scale of several millimeters. The BZ reaction can generate up to
several thousand oscillatory cycles in a closed system, which permits
studying chemical waves and patterns without constant replenishment of
reactants \cite{Zhabotinsky}.

Consider the water--in--oil micro--emulsion BZ reaction \cite{kve,kve2}
\begin{eqnarray}
\partial _{t}v &=&D_{v}\Delta v+\frac{1}{\varepsilon _{0}}\left[
f_{0}z+i_{0}\left( 1-mz\right) \right] \frac{v-q_{0}}{v+q_{0}}+\frac{1}{%
\varepsilon _{0}}\left[ \frac{1-mz}{1-mz+\varepsilon _{1}}\right] v-v^{2},
\notag \\
\partial _{t}z &=&D_{z}\Delta z-z+v\left[ \frac{1-mz}{1-mz+\varepsilon _{1}}%
\right] ,  \label{BZ1}
\end{eqnarray}%
where $v,z$ are dimensionless concentrations of activator HBrO$_{2}$ and
oxidized catalyst $[Ru(bpy)_{3}]^{3+}$ respectively; $D_{v}$ and $D_{z}$ are
dimensionless diffusion coefficients of activator and catalyst; $%
f,\varepsilon _{0},\varepsilon _{1}$ and $q$ are parameters of the standard
Tyson model \cite{Tyson}; $i_{0}$ represents the photoinduced production of
inhibitor, and $m$ represents the strength of oxidized state of the catalyst
with $0<mz<1$. This reaction was shown experimentally and numerically to
admit localized spot patterns that persist for long time \cite{kve,kve2}.

We can rescale the variables in (\ref{BZ1}) as \cite{Kolokolnikov}
\begin{equation*}
z=1/m-m^{-3/2}w\varepsilon _{1},\qquad v=m^{-1/2}\hat{v},\qquad
t=\varepsilon _{0}m^{1/2}\hat{t}.
\end{equation*}%
In the new variables, after dropping the hats, we obtain the
non--dimensional 2--component BZ reaction
\begin{equation*}
\partial _{t}v=\varepsilon ^{2}\Delta v+f(v,w),\qquad \tau \partial
_{t}w=D\Delta w+g(v,w),
\end{equation*}%
where
\begin{equation*}
f(v,z)=-\left[ f_{0}+f_{1}w\right] \frac{v-q}{v+q}+\left[ \frac{w}{1+\alpha w%
}\right] v-v^{2},\qquad g(v,w)=1-\left[ \frac{w}{1+\alpha w}\right] v,
\end{equation*}%
with the non-dimensional constants given by
\begin{eqnarray*}
&&\alpha =m^{-1/2},\qquad f_{1}=\varepsilon _{1}m^{1/2}\left( i_{0}-\frac{%
f_{0}}{m}\right) ,\qquad q=q_{0}m^{1/2}, \\
&&\varepsilon ^{2}=\varepsilon _{0}D_{v}m^{1/2},\qquad D=D_{z}\varepsilon
_{1}m^{-1/2},\qquad \tau =\frac{1}{m}\frac{\varepsilon _{1}}{\varepsilon _{0}%
}.
\end{eqnarray*}%

\subsection{3--component and multi--component systems}

\subsubsection{Oregonator}

The Oregonator model is based on the so--called FKN--mechanism \cite{FKNorig}%
, which provided the first successful explanation of the chemical
oscillations that occur in the experimental Belousov--Zhabotinsky reaction.
It is is composed of five coupled elementary chemical stoichiometries.
During the last two decades, the Oregonator model has been modified in many
ways by inclusion of additional chemical reaction steps or by changing the
rate constants. If we denote the concentration of the species S by [S], then
we define: $A=[$BrO$_{3}^{-}$], $H=[$H$^{+}$], $X=[$HBrO$_{2}$], $Y=[$Br$%
^{-} $], $Z=[$Ce$^{4+}$]. The original Oregonator model was described by the
following three coupled nonlinear PDEs,
\begin{eqnarray}
\partial _{t}X &=&k_{1}AH^{2}Y-k_{2}HXY-2k_{3}X^{2}+k_{4}AHX+D_{X}\nabla _{%
\mathbf{r}}^{2}X,  \notag \\
\partial _{t}Y &=&-k_{1}AH^{2}Y-k_{2}HXY+k_{5}fZ+D_{Y}\nabla _{\mathbf{r}%
}^{2}Y,  \label{Oreg} \\
\partial _{t}Z &=&2k_{4}AHX-k_{5}Z+D_{Z}\nabla _{\mathbf{r}}^{2}Z,  \notag
\end{eqnarray}%
where $f$ is a stoichiometric factor \cite{HopfQuench}, $k_{i}(i=1,...,5)$
are rate constants, while $D_{X}$, $D_{Y}$, and $D_{Z}$ are the diffusion
constants of the species HBrO$_{2}$, Br$^{-}$, and Ce$^{4+}$ respectively
(for dilute solutions, the diffusion matrix is diagonal). For a thorough
discussion of the chemistry on which the Oregonator is based, the reader is
referred to \cite{Tyson}.

The Oregonator temporal mass--action dynamics is a well--stirred,
homogeneous system of ODEs given by
\begin{eqnarray*}
\dot{X} =k_{1}AY-k_{2}XY+k_{3}AX-2k_{4}X^{2},\qquad \dot{Y}%
=-k_{1}AY-k_{2}XY+1/2k_{c}fBZ,\qquad  \\
\dot{Z} =2k_{3}AX-k_{c}BZ,\hspace{5cm}
\end{eqnarray*}%
which are typically scaled as \cite{Tyson}%
\begin{equation*}
\epsilon (dx/d\tau )=qy-xy+x(1-x),\qquad \epsilon ^{\prime }(dy/d\tau
)=-qy-xy+fz,\qquad dz/d\tau =x-z.
\end{equation*}%
The basic chemistry of the BZ--oscillations involves jumps between high and
low HBrO2 ($X$) states, which is reflected in the relaxation oscillator
nature of the Oregonator. This fundamental bistability may be stabilized in
a flow reactor (CSTR) with reactants and Br$^{-}$ in the feed stream.
Hysteresis between the two states is observed both experimentally and in the
Oregonator. Quasiperiodicity and chaos also are observed in CSTR and can be
modeled by the Oregonator \cite{Field07}.

\subsubsection{Multi--phase tumor growth equations}

Our last reaction--diffusion system is a general model of
multi--phase tumor growth, in the form of nonlinear parabolic PDE,
as reviewed recently in \cite{Roose}
\begin{equation}
\partial _{t}\Phi _{i}=\nabla \cdot (D_{i}\Phi _{i})-\nabla \cdot (\mathbf{v}%
_{i}\Phi _{i})+\lambda _{i}(\Phi _{i},C_{i})-\mu _{i}(\Phi _{i},C_{i})
\label{multiPh}
\end{equation}
($\partial _{t}\equiv \partial /\partial {t}$), where for phase $i$, $\Phi
_{i}$ is the volume fraction ($\sum_{i}\Phi _{i}=1 $), $D_{i}$ is the random
motility or diffusion, $\lambda _{i}(\Phi _{i},C_{i})$ is the chemical and
phase dependent production, and $\mu _{i}(\Phi _{i},C_{i})$ is the chemical
and phase dependent degradation/death, and $\mathbf{v}_{i}$ is the cell
velocity defined by the constitutive equation
\begin{equation}
\mathbf{v}_{i}=-\mu \nabla p,  \label{constit}
\end{equation}
where $\mu $ is a positive constant describing the viscous--like properties
of tumor cells and $p$ is the spheroid internal pressure.

In particular, the multi--phase equation (\ref{multiPh}) splits into two
heat--like mass--conservation PDEs \cite{Roose},
\begin{equation}
\partial _{t}\Phi ^{C}=S^{C}-\nabla \cdot (\mathbf{v}^{C}\Phi ^{C}),\qquad
\partial _{t}\Phi ^{F}=S^{F}-\nabla \cdot (\mathbf{v}^{F}\Phi ^{F}),
\label{2Ph}
\end{equation}
where $\Phi ^{C}$ and $\Phi ^{F}$ are the tissue cell/matrix and fluid
volume fractions, respectively, $\mathbf{v}^{C}$ and $\mathbf{v}^{F}$ are
the cell/matrix and the fluid velocities (both defined by their constitutive
equations of the form of (\ref{constit})), $S^{C}$ is the rate of production
of solid phase tumor tissue and $S^{F}$ is the creation/degradation of the
fluid phase. Conservation of matter in the tissue, $\Phi ^{C}+\Phi ^{F}=1$,
implies that $\nabla \cdot $($\mathbf{v}^{C}\Phi ^{C}$ +$\mathbf{v}^{F}\Phi
^{F}$) = $\Phi ^{C}+\Phi ^{F}$. The assumption that the tumor may be
described by two phases only implies that the new cell/matrix phase is
formed from the fluid phase and vice versa, so that $S^{C}+S^{F}=0$. The
detailed biochemistry of tumor growth can be coupled into the model above
through the growth term $S^{C}$, with equations added for nutrient
diffusion, see \cite{Roose} and references therein.

The multi--phase tumor growth model (\ref{multiPh}) has been derived from
the classical transport/mass conservation equations for different chemical
species \cite{Roose},
\begin{equation}
\partial _{t}u_{i}=P_{i}-\nabla \cdot \mathbf{N}_{i}.  \label{conEq}
\end{equation}%
Here $C_{i}$ are the concentrations of the chemical species, subindex $a$
for oxygen, $b$ for glucose, $c$ for lactate ion, $d$ for carbon dioxide, $e$
for bicarbonate ion, $f$ for chloride ion, and $g$ for hydrogen ion
concentration; $P_{i}$ is the net rate of consumption/production of the
chemical species both by tumor cells and due to the chemical reactions with
other species; and $\mathbf{N}_{i}$ is the flux of each of the chemical
species inside the tumor spheroid, given (in the simplest case of uncharged
molecules of glucose, $O_{2}$ and $CO_{2}$) by Fick's law,
\begin{equation*}
\mathbf{N}_{i}=-D_{i}\nabla u_{i},
\end{equation*}%
where $D_{i}$ are (positive) constant diffusion coefficients. In case of
charged molecules of ionic species, the flux $\mathbf{N}_{i}$ contains also
the (negative) gradient of the volume fractions $\Phi _{i}$.

There are three distinct stages to cancer development: avascular, vascular,
and metastatic -- researchers often concentrate their efforts on answering
specific OUPC--related questions on each of these stages \cite{Roose}. In
particular, as some tumor cell lines grown in vitro form spherical
aggregates, the relative cheapness and ease of in vitro experiments in
comparison to animal experiments has made 3D \emph{multicellular tumor
spheroids} (MTS, see Figure 6 in \cite{Roose}) very popular in vitro model
system of avascular tumors\footnote{%
In vitro cultivation of tumor cells as multicellular tumor spheroids (MTS)
has greatly contributed to the understanding of the role of the cellular
micro-environment in tumor biology (for review see \cite{Sutherland,Kunz}).
These spherical cell aggregates mimic avascular tumor stages or
micro-metastases in many aspects and have been studied intensively as an
experimental model reflecting an in vivo-like micro-milieu with 3D metabolic
gradients. With increasing size, most MCTS not only exhibit proliferation
gradients from the periphery towards the center but they also develop a
spheroid type-specific nutrient supply pattern, such as radial oxygen
partial pressure gradients. Similarly, MCTS of a variety of tumor cell lines
exhibit a concentric histo-morphology, with a necrotic core surrounded by a
viable cell rim. The spherical symmetry is an important prerequisite for
investigating the effect of environmental factors on cell proliferation and
viability in a 3D environment on a quantitative basis.} \cite{Kunz}. They
are used to study how local micro-environments affect cellular growth/decay,
viability, and therapeutic response \cite{Sutherland}. MTS provide, allowing
strictly controlled nutritional and mechanical conditions, excellent
experimental patterns to test the validity of the proposed mathematical
models of tumor growth/decay \cite{Preziosi}.

\section{Dissipative evolution under the Ricci flow}

In this section we will derive the geometric formalism associated
with the quadratic Ricci--flow equation (\ref{RH}), as a general
framework for all presented bio--reaction--diffusion systems.

\subsection{Geometrization Conjecture}

Geometry and topology of smooth surfaces are related by the \textit{%
Gauss--Bonnet formula} for a closed surface $\Sigma $ (see, e.g., \cite%
{GaneshSprBig,GaneshADG})
\begin{equation}
\frac{1}{2\pi}\iint_{\Sigma }K\,dA= \chi (\Sigma )=2-2\,\mathrm{gen}(\Sigma
),  \label{GB}
\end{equation}
where $dA$ is the area element of a metric $g$ on $\Sigma $, $K$ is the
Gaussian curvature, $\chi (\Sigma )$ is the Euler characteristic of $\Sigma $
and $\mathrm{gen}(\Sigma )$ is its \emph{genus}, or number of handles, of $%
\Sigma $. Every closed surface $\Sigma $ admits a metric of constant
Gaussian curvature $K=+1,\,0$, or $-1$ and so is uniformized by elliptic,
Euclidean, or hyperbolic geometry, which respectively have $\mathrm{gen}%
(S^2)=0$ (sphere), $\mathrm{gen}(T^2)=1$ (torus) and $\mathrm{gen}(\Sigma
)>1 $ (torus with several holes). The integral (\ref{GB}) is a \emph{%
topological invariant} of the surface $\Sigma $, always equal to 2
for all topological spheres $S^2$ (that is, for all closed
surfaces without holes that can be continuously deformed from the
geometric sphere) and always equal to 0 for the topological torus
$T^2$ (i.e., for all closed surfaces with one hole or handle).

Topological framework for the Ricci flow (\ref{RF}) is Thurston's
\textit{Geometrization Conjecture} \cite{Thurston}, which states
that the interior of any compact 3--manifold can be split in an
essentially unique way by disjoint embedded 2--spheres $S^{2}$ and
tori $T^{2}$ into pieces and each piece admits one of 8 geometric
structures (including (i) the 3--sphere $S^{3}$ with constant
curvature $+1$; (ii) the Euclidean 3--space $\mathbb{R}^{3}$ with
constant curvature 0 and (iii) the hyperbolic
3--space $\mathbb{H}^{3}$ with constant curvature $-1$).\footnote{%
Another five allowed geometric structures are represented by the following
examples: (iv) the product $S^{2}\times S^{1}$; (v) the product $\mathbb{H}%
^{2}\times S^{1}$ of hyperbolic plane and circle; (vi) a left invariant
Riemannian metric on the special linear group $SL(2,\mathbb{R})$; (vii) a
left invariant Riemannian metric on the solvable Poincar\'{e}-Lorentz group $%
E(1,1)$, which consists of rigid motions of a ($1+1)-$dimensional space-time
provided with the flat metric $dt^{2}-dx^{2}$; (viii) a left invariant
metric on the nilpotent Heisenberg group, consisting of $3\times 3$ matrices
of the form
\par
$\left[
\begin{array}{ccc}
1 & \ast & \ast \\
0 & 1 & \ast \\
0 & 0 & 1%
\end{array}%
\right] .$ In each case, the universal covering of the indicated manifold
provides a canonical model for the corresponding geometry \cite{Milnor}.}
The geometrization conjecture (which has the Poincar\'{e} Conjecture as a
special case) would give us a link between the geometry and topology of
3--manifolds, analogous in spirit to the case of 2--surfaces.

In higher dimensions, the Gaussian curvature $K$ corresponds to the Riemann
curvature tensor $\mathfrak{Rm}$ on a smooth $n-$manifold $M$, which is in
local coordinates on $M$ denoted by its $(4,0)-$components $R_{ijkl}$, or
its $(3,1)-$components $R_{ijk}^{l}$ (see Appendix, as well as e.g., \cite%
{GaneshSprBig,GaneshADG}). The trace (or, contraction) of $\mathfrak{Rm}$,
in $(4,0)-$case using the inverse metric tensor $g^{ij}=(g_{ij})^{-1}$, is the Ricci tensor $%
\mathfrak{Rc}$, the contracted curvature tensor, which is in a local
coordinate system $\{x^{i}\}_{i=1}^{n}$ defined in an open set $U\subset M$,
given by
\begin{equation*}
R_{ij}=\mathrm{tr}(\mathfrak{Rm})=g^{kl}R_{ijkl}
\end{equation*}%
(using Einstein's summation convention), while the \textit{scalar curvature}
is now given by the second contraction of $\mathfrak{Rm}$ as
\begin{equation*}
R=\mathrm{tr}(\mathfrak{Rc})=g^{ij}R_{ij}.
\end{equation*}

In general, the Ricci flow $g_{ij}(t)$ is a one--parameter family
of Riemannian metrics on a compact $n-$manifold $M$ governed by
the equation (\ref{RF}), which has a unique solution for a short time for an arbitrary smooth metric $%
g_{ij}$ on $M$ \cite{Ham82}. If $\mathfrak{Rc}>0$ at any local point $%
x=\{x^i\}$ on $M$, then the Ricci flow (\ref{RF}) contracts the metric $%
g_{ij}(t)$ near $x$, to the future, while if $\mathfrak{Rc}<0$, then the
flow (\ref{RF}) expands $g_{ij}(t)$ near $x$. The solution metric $g_{ij}(t)$
of the Ricci flow equation (\ref{RF}) shrinks in positive Ricci curvature
direction while it expands in the negative Ricci curvature direction,
because of the minus sign in the front of the Ricci tensor $R_{ij} $. In
particular, on a 2--sphere $S^{2}$, any metric of positive Gaussian
curvature will shrink to a point in finite time. At a general point, there
will be directions of positive and negative Ricci curvature along which the
metric will locally contract or expand (see \cite{Anderson}). Also, if a
simply--connected compact 3--manifold $M$ has a Riemannian metric $g_{ij}$
with positive Ricci curvature then it is diffeomorphic to the 3--sphere $S^3$
\cite{Ham82}. More generally speaking, the Ricci flow deforms manifolds with positive Ricci curvature to a point which can be renormalized to the 3--sphere.

\subsection{Reaction--diffusion--type evolution of curvatures and volumes}

All three Riemannian curvatures ($R,\mathfrak{Rc}$ and $\mathfrak{Rm}$), as
well as the associated volume forms, \emph{evolve} during the Ricci flow (%
\ref{RF}). In general, the Ricci--flow evolution equation
(\ref{RF}) for the metric tensor $g_{ij}$ implies the
reaction--diffusion--type evolution equation for the Riemann
curvature tensor $\mathfrak{Rm}$ on an $n-$manifold $M$,
\begin{equation}
\partial _{t}\mathfrak{Rm}=\Delta \mathfrak{Rm}+Q_n,  \label{RMflow}
\end{equation}%
where $Q_n$ is a quadratic expression of the Riemann
$n-$curvatures, corresponding to the $n-$component bio--chemical
reaction, while the term $\Delta \mathfrak{Rm}$ corresponds to the
$n-$component diffusion. From the general $n-$curvature evolution
(\ref{RMflow}) we have two important
particular cases:\footnote{%
By expanding the maximum principle for tensors, Hamilton proved that Ricci
flow $g(t)$ given by (\ref{RF}) preserves the positivity of the Ricci tensor
$\mathfrak{Rc}$ on 3--manifolds (as well as of the Riemann curvature tensor $%
\mathfrak{Rm}$ in all dimensions); moreover, the eigenvalues of the Ricci
tensor on 3--manifolds (and of the curvature operator $\mathfrak{Rm}$ on
4--manifolds) are getting pinched point-wisely as the curvature is getting
large \cite{Ham82,4-manifold}. This observation allowed him to prove the
convergence results: the evolving metrics (on a compact manifold) of
positive Ricci curvature in dimension 3 (or positive Riemann curvature in
dimension 4) converge, modulo scaling, to metrics of constant positive
curvature.
\par
However, without assumptions on curvature, the long time behavior of the
metric evolving by Ricci flow may be more complicated \cite{Perel1}. In
particular, as $t$ approaches some finite time $T$, the curvatures may
become arbitrarily large in some region while staying bounded in its
complement. On the other hand, Hamilton \cite{Harnack} discovered a
remarkable property of solutions with nonnegative curvature tensor $%
\mathfrak{Rm}$ in arbitrary dimension, called the \emph{differential Harnack
inequality}, which allows, in particular, to compare the curvatures of the
solution of (\ref{RF}) at different points and different times.}

1. The evolution equation for the Ricci curvature tensor
$\mathfrak{Rc}$ on a 3--manifold $M$,
\begin{equation}
\partial _{t}\mathfrak{Rc}=\Delta \mathfrak{Rc}+Q_{3},  \label{Rcflow}
\end{equation}%
where $Q_{3}$ is a quadratic expression of the Ricci 3--curvatures,
corresponding to the 3--component bio--chemical reaction, while the term $%
\Delta \mathfrak{Rc}$ corresponds to the 3--diffusion.

2. The evolution equation for the scalar curvature $R$,
\begin{equation}
\partial _{t}R=\Delta R+2|\mathfrak{Rc}|^{2},  \label{DTR}
\end{equation}%
in which the term $2|\mathfrak{Rc}|^{2}$ corresponds to the
2--component bio--chemical reaction, while the term $\Delta R$
corresponds to the 2--component diffusion. By the \textit{maximum
principle} (see subsection \ref{Rentr}), the minimum of the scalar
curvature $R$ is non--decreasing along the flow $g(t)$, both on
$M$ and on its boundary $\partial M$ (see \cite{Perel1}).

Let us now see in detail how various geometric quantities evolve
given the \emph{short--time solution} of the Ricci flow equation
(\ref{RF}) on an arbitrary $n-$manifold $M$. For this, we need
first to calculate the \textit{variation formulas} for the
Christoffel symbols and curvature tensors on $M$; then the
corresponding evolution equations will naturally follow (see \cite%
{Ham82,CaoChow,ChowKnopf}). If $g(s)$ is a 1--parameter family of
metrics on $M$ with
$
\partial _{s}g_{ij}=v_{ij},
$
then the variation of the Christoffel symbols $\Gamma _{ij}^{k}$ on $M$ is
given by
\begin{equation}
\partial _{s}\Gamma _{ij}^{k}=\frac{1}{2}g^{kl}\left( \nabla
_{i}v_{jl}+\nabla _{j}v_{il}-\nabla _{l}v_{ij}\right) ,  \label{VC}
\end{equation}%
(where $\nabla$ is the covariant derivative with respect to the Riemannian connection) from which follows the evolution of the Christoffel symbols $\Gamma
_{ij}^{k} $ under the Ricci flow $g(t)$ on $M$ given by (\ref{RF}),%
\begin{equation*}
\partial _{t}\Gamma _{ij}^{k}=-g^{kl}\left( \nabla _{i}R_{jl}+\nabla
_{j}R_{il}-\nabla _{l}R_{ij}\right) .
\end{equation*}

From (\ref{VC}) we calculate the variation of the Ricci tensor $R_{ij}$ on $%
M $ as
\begin{equation}
\partial _{s}R_{ij}=\nabla _{m}\left( \partial _{s}\Gamma _{ij}^{m}\right)
-\nabla _{i}\left( \partial _{s}\Gamma _{mj}^{m}\right) ,  \label{VRic}
\end{equation}%
and the variation of the scalar curvature $R$ on $M$ by
\begin{equation}
\partial _{s}R=-\Delta V+\mathrm{div}(\mathrm{div\,}v)-\left\langle v,%
\mathfrak{Rc}\right\rangle ,  \label{VR}
\end{equation}%
where $V=g^{ij}v_{ij}=\mathrm{tr}(v)$ is the trace of $v=(v_{ij})$.

If an $n-$manifold $M$ is oriented, then the \textit{volume}
$n-$form on $M$ is given, in a positively--oriented local
coordinate system $\{x^{i}\}\in U\subset M$, by \cite{VladSiam}
\begin{equation}
d\mu =\sqrt{\det (g_{ij})}\,dx^{1}\wedge dx^{2}\wedge...\wedge dx^{n}.
\label{dmu}
\end{equation}%
If $\partial _{s}g_{ij}=v_{ij},$ then
$
\partial _{s}d\mu =\frac{1}{2}Vd\mu .
$
The evolution of the volume $n-$form $d\mu $ under the Ricci flow $g(t)$ on $%
M$ is given by the exponential decay/growth relation with the scalar
curvature $R$ as the (variable) rate parameter,
\begin{equation}
\partial _{t}d\mu =-Rd\mu,  \label{dtmu}
\end{equation}
which gives an exponential decay for $R>=a>0$ (elliptic geometry) and
exponential growth for $R<=a<0$ (hyperbolic geometry) -- for any small constant $a$ (scalar curvature must be bounded away from zero). The elementary volume
evolution (\ref{dtmu}) implies the integral form of the exponential relation
for the total $n-$volume
\begin{equation*}
\mathrm{vol}(g)=\int_{M}d\mu ,\qquad \text{as\qquad }\partial _{t}\mathrm{vol%
}(g(t))=-\int_{M}Rd\mu ,
\end{equation*}
which again gives an \emph{exponential decay} for elliptic $R>0$ and \emph{%
exponential growth} for hyperbolic $R<0$.

This is a crucial point for the \emph{tumor suppression} by the body: the
immune system needs to keep the elliptic geometry of the MTS, evolving by (%
\ref{multiPh}) -- by all possible means.\footnote{%
As a tumor decay control tool, a monoclonal antibody therapy is usually
proposed. Monoclonal antibodies (mAb) are mono-specific antibodies that are
identical because they are produced by one type of immune cell that are all
clones of a single parent cell. Given (almost) any substance, it is possible
to create monoclonal antibodies that specifically bind to that substance;
they can then serve to detect or purify that substance. The invention of
monoclonal antibodies is generally accredited to Georges K\"{o}hler, C\'{e}%
sar Milstein, and Niels Kaj Jerne in 1975 \cite{Kohler}, who shared the
Nobel Prize in Physiology or Medicine in 1984 for the discovery. The key
idea was to use a line of myeloma cells that had lost their ability to
secrete antibodies, come up with a technique to fuse these cells with
healthy antibody producing B--cells, and be able to select for the
successfully fused cells.} In the healthy organism this normally happens,
because the initial MTS started as a spherical shape with $R>0$. The immune
system just needs to keep the MTS in the spherical/elliptic shape and
prevent any hyperbolic distortions of $R<0$. Thus, it will naturally have an
exponential decay and vanish.

Since the $n-$volume is not constant and sometimes we would like to prevent
the solution from shrinking to an $n-$point on $M$ (elliptic case) or
expanding to an infinity (hyperbolic case), we can also consider the \textit{%
normalized Ricci flow} on $M$ (see \cite{CaoChow}):
\begin{equation}
\partial _{t}\hat{g}_{ij}=-2\hat{R}_{ij}+\frac{2}{n}\hat{r}\hat{g}%
_{ij},\qquad \text{where\qquad }\hat{r}=\mathrm{vol}(\hat{g})^{-1}\int_{M}%
\hat{R}d\mu   \label{NRF}
\end{equation}%
is the average scalar curvature on $M$. We then have the
$n-$\emph{volume conservation law:}
\begin{equation*}
\partial _{t}\mathrm{vol}(\hat{g}(t))=0.
\end{equation*}

To study the \emph{long--time existence} of the normalized Ricci flow (\ref%
{NRF}) on an arbitrary $n-$manifold $M$, it is important to know what kind
of curvature conditions are preserved under the equation. In general, the
Ricci flow $g(t) $ on $M$, as defined by the fundamental relation (\ref{RF}%
), tends to preserve some kind of positivity of curvatures. For example,
positive scalar curvature $R$ (i.e., elliptic geometry) is preserved both on
$M$ and on its boundary $\partial M$ in any dimension. This follows from
applying the maximum principle to the evolution equation (\ref{DTR}) for
scalar curvature $R$ both on $M$ and on $\partial M$. Similarly, positive
Ricci curvature is preserved under the Ricci flow on a 3--manifold $M$. This
is a special feature of dimension 3 and is related to the fact that the
Riemann curvature tensor may be recovered algebraically from the Ricci
tensor and the metric on 3--manifolds \cite{CaoChow}.

In particular, we have the following result for 2--surfaces (see
\cite{surface}): Let $S=\partial M$ be a closed 2--surface, which
is a boundary of a compact 3--manifold $M$. Then for any initial
2--metric $g_{0}$ on $\partial M$, the solution to the normalized
Ricci flow (\ref{NRF}) on $\partial M$ exists for all time. In other words, the normalized Ricci flow in 2D always converges.
Moreover, (i) if the Euler characteristic of $\partial M$ is
non--positive, then the solution metric $g(t)$ on $\partial M$
converges to a constant curvature metric as $t\rightarrow \infty
$; and (ii) if the scalar curvature
$R$ of the initial metric $g_{0}$ is positive, then the solution metric $%
g(t) $ on $\partial M$ converges to a positive constant curvature metric as $%
t\rightarrow \infty .$ (For surfaces with non--positive Euler
characteristic, the proof is based primarily on maximum principle estimates
for the scalar curvature.)

Applying to the tumor evolution (\ref{multiPh}), the normalized Ricci flow (%
\ref{NRF}) of the MTS will make it completely round with a
geometric sphere shell, which is ideal for surgical removal. This
is our second option for the MTS growth/decay control. If we
cannot force it to exponential decay, then we must try to
normalize into a round spherical shell -- which is suitable for
surgical removal.

The negative flow of the total $n-$volume $\mathrm{vol}(g(t))$ represents
the \textit{Einstein--Hilbert functional} (see \cite{MTW,CaoChow,Anderson})
\begin{equation*}
E(g)=\int_{M}Rd\mu =-\partial _{t}\mathrm{vol}(g(t)).
\end{equation*}%
If we put $\partial _{s}g_{ij}=v_{ij},$ we have%
\begin{equation*}
\partial _{s}E(g)=\int_{M}\left( -\Delta V+\mathrm{div}(\mathrm{div\,}%
v)-\left\langle v,\mathfrak{Rc}\right\rangle +\frac{1}{2}RV\right)
d\mu =\int_{M}\left\langle
v,\frac{1}{2}Rg_{ij}-R_{ij}\right\rangle d\mu ,
\end{equation*}
so, the critical points of $E(g)$ satisfy \emph{Einstein's equation }$\frac{1%
}{2}Rg_{ij}=R_{ij}$ in the vacuum. The gradient flow of $E(g)$ on an $n-$%
manifold $M$,
\begin{equation*}
\partial _{t}g_{ij}=2\left( \nabla E(g)\right) _{ij}=Rg_{ij}-2R_{ij},
\end{equation*}%
is the Ricci flow (\ref{RF}) plus $Rg_{ij}$. Thus, Einstein metrics are the
fixed points of the normalized Ricci flow.\footnote{%
Einstein metrics on $n-$manifolds are metrics with constant Ricci curvature.
However, along the way, the deformation will encounter singularities. The
major question, resolved by Perelman, was how to find a way to describe all
possible singularities.}

Let $\Delta $ denote the Laplacian acting on functions on a closed $n-$%
manifold $M$, which is in local coordinates $\{x^{i}\}\in U\subset M$ given
by%
\begin{equation*}
\Delta =g^{ij}\nabla _{i}\nabla _{j}=g^{ij}\left( \partial _{ij}-\Gamma
_{ij}^{k}\partial _{k}\right) .
\end{equation*}%
For any smooth function $f$ on $M$ we have \cite{Ham82,ChowKnopf}%
\begin{equation*}
\Delta \nabla _{i}f=\nabla _{i}\Delta f+R_{ij}\nabla _{j}f\quad
\text{and\quad}\Delta |\nabla f|^{2}=2|\nabla _{i}\nabla
_{j}f|^{2}+2R_{ij}\nabla _{i}f\nabla _{j}f+2\nabla _{i}f\nabla
_{i}\Delta f,
\end{equation*}
from which it follows that if we have
\begin{eqnarray*}
\mathfrak{Rc} &\geq &0,\qquad \Delta f\equiv 0,\qquad |\nabla f|\equiv
1,\qquad \text{then} \\
\nabla \nabla f &\equiv &0\qquad \text{and\qquad }\mathfrak{Rc}(\nabla
f,\nabla f)\equiv 0.
\end{eqnarray*}

Using this Laplacian $\Delta $, we can write the linear heat equation on $M$
as \ $\partial _{t}u=\Delta u,$ where $u$ is the temperature. In particular,
the Laplacian acting on functions with respect to $g(t)$ will be denoted by $%
\Delta _{g(t)}$. If $(M,g(t))$ is a solution to the Ricci flow equation (\ref%
{RF}), then we have%
\begin{equation*}
\partial _{t}\Delta _{g(t)}=2R_{ij}\nabla _{i}\nabla _{j}.
\end{equation*}

The evolution equation (\ref{DTR}) for the scalar
curvature $R$ under the Ricci flow (\ref{RF}) follows from (\ref{VR}).
Using equation (\ref{cB}) from Appendix, we have:
\begin{equation*}
\mathrm{div}(\mathfrak{Rc})=\frac{1}{2}\nabla R,\quad\text{so that}\quad
\mathrm{div}(\mathrm{div}(\mathfrak{Rc})) = \frac{1}{2}\Delta R,
\end{equation*}
showing again that the scalar curvature $R$ satisfies a heat--type equation
with a quadratic nonlinearity both on a 3--manifold $M$ and on its boundary
2--surface $\partial M$.

Next we will find the exact form of the evolution equation (\ref{Rcflow})
for the Ricci tensor $\mathfrak{Rc}$ under the Ricci flow $g(t)$ given by (%
\ref{RF}) on any $3-$manifold $M$. (Note that in higher
dimensions, the appropriate formula of huge complexity would
involve the whole Riemann curvature tensor $\mathfrak{Rm}$.) Given
a variation $\partial _{s}g_{ij}=v_{ij}$, from (\ref{VRic}) we get
\begin{equation*}
\partial _{s}R_{ij}=\frac{1}{2}\left( \Delta _{L}v_{ij}+\nabla _{i}\nabla
_{j}V-\nabla _{i}(\mathrm{div\,}v)_{j}-\nabla _{j}(\mathrm{div\,}%
v)_{i}\right) ,
\end{equation*}%
where $\Delta _{L}$ denotes the so--called Lichnerowicz Laplacian (which
depends on $\mathfrak{Rm}$, see \cite{Ham82,ChowKnopf}). Since%
\begin{equation*}
\nabla _{i}\nabla _{j}R-\nabla _{i}(\mathrm{div}(\mathfrak{Rc}))_{j}-\nabla
_{j}(\mathrm{div}(\mathfrak{Rc}))_{i}=0,
\end{equation*}%
by (\ref{cB}) (after some algebra) we get that under the Ricci flow (\ref{RF}%
) the evolution equation for the Ricci tensor $\mathfrak{Rc}$ on a
3--manifold $M$ is
\begin{equation*}
\partial _{t}R_{ij}=\Delta R_{ij}+3RR_{ij}-6R_{im}R_{jm}+\left( 2|\mathfrak{%
Rc}|^{2}-R^{2}\right) g_{ij}.
\end{equation*}%
So, just as in case of the evolution (\ref{DTR}) of the scalar curvature $%
\partial _{t}R$ (both on a 3--manifold $M$ and on its 2--boundary $\partial
M $), we get a heat--type evolution equation with a quadratic nonlinearity
for $\partial _{t}R_{ij}$, which means that positive Ricci curvature ($%
\mathfrak{Rc}>0$) of elliptic 3--geometry is preserved under the Ricci flow $%
g(t)$ on $M$.

More generally, we have the following result for 3--manifolds (see \cite%
{Ham82}): Let $(M,g_{0})$ be a compact Riemannian $3-$manifold with positive
Ricci curvature $\mathfrak{Rc}$. Then there exists a unique solution to the
normalized Ricci flow $g(t)$ on $M$ with $g(0)=g_{0}$ for all time and the
metrics $g(t)$ converge exponentially fast to a constant positive sectional
curvature metric $g_{\infty }$ on $M$. In particular, $M$ is diffeomorphic
to a 3--sphere $S^3$. (As a consequence, such a $3-$manifold $M$ is
necessarily diffeomorphic to a quotient of the $3-$sphere by a finite group
of isometries. It follows that given any homotopy $3-$sphere, if one can
show that it admits a metric with positive Ricci curvature, then the Poincar%
\'{e} Conjecture would follow \cite{CaoChow}.) In addition, compact and
closed $3-$manifolds which admit a non-singular solution can also be
decomposed into geometric pieces \cite{non-singular}.

From the geometric evolution equations reviewed in this
subsection, we see that both short--time and long--time geometric
solution can always be found for 2--compo-nent
bio--reaction--diffusion equations, as they correspond to
evolution of the scalar 2--curvature $R$. Regarding the
3--component bio--reaction--diffusion equations, corresponding to
evolution of the Ricci 3--curvature $\mathfrak{Rc},$ we can always
find the short--time geometric solution, while the long--time
solution exists only under some additional (compactnes and/or
closure) conditions. Finally, in case of $n-$component
bio--reaction--diffusion equations, corresponding to evolution of
the Riemann $n-$curvature $\mathfrak{Rm},$ only short--time
geometric solution is possible.

\subsection{Dissipative solitons and Ricci breathers}

An important class of bio--reaction--diffusion systems are
\emph{dissipative solitons} (DSs), which are stable solitary
localized structures that arise in nonlinear spatially extended
dissipative systems due to mechanisms of
\emph{self--organization}. They can be considered as an extension
of the classical soliton concept in conservative systems. Apart
from aspects similar to the behavior of classical particles like
the formation of bound states, DSs exhibit entirely nonclassical
behavior -- e.g., scattering, generation and annihilation -- all
without the constraints of energy or momentum conservation. The
excitation of internal degrees of freedom may result in a
dynamically stabilized intrinsic speed, or periodic oscillations
of the shape.

In particular, stationary DSs are generated by production of material in the
center of the DSs, diffusive transport into the tails and depletion of
material in the tails. A propagating pulse arises from production in the
leading and depletion in the trailing end. Among other effects, one finds
periodic oscillations of DSs, the so--called `breathing' dissipative
solitons \cite{Gurevich}.

DSs in many different systems show universal particle--like properties. To
understand and describe the latter, one may try to derive `particle
equations' for slowly varying order parameters like position, velocity or
amplitude of the DSs by adiabatically eliminating all fast variables in the
field description. This technique is known from linear systems, however
mathematical problems arise from the nonlinear models due to a coupling of
fast and slow modes \cite{Friedrich}.

Similar to low--dimensional dynamic systems, for supercritical bifurcations
of stationary DSs one finds characteristic normal forms essentially
depending on the symmetries of the system; e.g., for a transition from a
symmetric stationary to an intrinsically propagating DS one finds the
Pitchfork normal form for the DS--velocity $\mathbf{v}$ \cite{Bode},
\begin{equation*}
\mathbf{\dot{v}}=(\sigma -\sigma _{0})\mathbf{v}-|\mathbf{v}|^{2}\mathbf{v}
\end{equation*}%
where $\sigma $ represents the bifurcation parameter and $\sigma _{0}$ the
bifurcation point. For a bifurcation to a `breathing' DS, one finds the Hopf
normal form \cite{StrAttr}
\begin{equation*}
\dot{A}=(\sigma -\sigma _{0})A-|A|^{2}A
\end{equation*}%
for the amplitude $A$ of the oscillation. Note that the above problems do
not arise for classical solitons as inverse scattering theory yields
complete analytical solutions \cite{Gurevich}.

\subsubsection{Ricci breathers and Ricci solitons}

Closely related to dissipative solitons are the so--called
\emph{breathers}, solitonic structures given by localized periodic
solutions of some nonlinear soliton PDEs, including the exactly--solvable sine--Gordon equation\footnote{%
An exact solution $u=u(x,t)$ of the (1+1)D sine--Gordon equation
\begin{equation*}
\partial _{t^{2}}u=\partial _{x^{2}}u-\sin u,\quad \text{is }\cite{Ablowitz}
\end{equation*}%
\begin{equation*}
u=4\arctan \left( \frac{\sqrt{1-\omega ^{2}}\;\cos (\omega t)}{\omega
\;\cosh (\sqrt{1-\omega ^{2}}\;x)}\right) ,
\end{equation*}%
which, for $\omega <1$, is periodic in time $t$ and decays exponentially
when moving away from $x=0$.} and the focusing nonlinear Schr\"{o}dinger
equation.\footnote{%
The focusing nonlinear Schr\"{o}dinger equation is the dispersive
complex--valued (1+1)D PDE \cite{Akhmediev},
\begin{equation*}
i\,\partial _{t}u+\partial _{x^{2}}u+|u|^{2}u=0,
\end{equation*}%
with a breather solution of the form:
\begin{equation*}
u=\left( \frac{2\,b^{2}\cosh (\theta )+2\,i\,b\,\sqrt{2-b^{2}}\;\sinh
(\theta )}{2\,\cosh (\theta )-\sqrt{2}\,\sqrt{2-b^{2}}\cos (a\,b\,x)}%
-1\right) \;a\;\exp (i\,a^{2}\,t)\quad \text{with}\quad \theta =a^{2}\,b\,%
\sqrt{2-b^{2}}\;t,
\end{equation*}%
which gives breathers periodic in space $x$ and approaching the uniform
value $a$ when moving away from the focus time $t=0$.}

A metric $g_{ij}(t)$ evolving by the Ricci flow $g(t)$ given by (\ref{RF})
on any 3--manifold $M$ is called a \emph{Ricci breather}, if for some $%
t_{1}<t_{2}$ and $\alpha >0$ the metrics $\alpha g_{ij}(t_{1})$ and $%
g_{ij}(t_{2})$ differ only by a diffeomorphism; the cases $\alpha =1,\alpha
<1,\alpha >1$ correspond to steady, shrinking and expanding breathers,
respectively. Trivial breathers on $M$, for which the metrics $g_{ij}(t_{1})$
and $g_{ij}(t_{2})$ differ only by diffeomorphism and scaling for each pair
of $t_{1}$ and $t_{2}$, are called \textit{Ricci solitons}. Thus, if one
considers Ricci flow as a dynamical system on the space of Riemannian
metrics modulo diffeomorphism and scaling, then breathers and solitons
correspond to periodic orbits and fixed points respectively. At each time
the Ricci soliton metric satisfies on $M$ an equation of the form \cite%
{Perel1}
\begin{equation*}
R_{ij}+cg_{ij}+\nabla _{i}b_{j}+\nabla _{j}b_{i}=0,
\end{equation*}%
where $c$ is a number and $b_{i}$ is a 1--form; in particular, when $b_{i}=%
\frac{1}{2}\nabla _{i}a$ for some function $a$ on $M,$ we get a gradient
Ricci soliton. An important example of a gradient shrinking soliton is the
\textit{Gaussian soliton}, for which the metric $g_{ij}$ is just the
Euclidean metric on $\mathbb{R}^{3}$, $c=1$ and $a=-|x|^{2}/2$.

\subsection{Smoothing/Averaging heat equation and Ricci entropy}
\label{Rentr}

Given a $C^{2}$ function $u:M\rightarrow \mathbb{R}$ on a Riemannian $3-$%
manifold $M$, its Laplacian is defined in local coordinates $\left\{
x^{i}\right\}\in U\subset M $ to be
\begin{equation*}
\Delta u=\text{\textrm{tr}}_{g}\left( \nabla ^{2}u\right) =g^{ij}\nabla
_{i}\nabla _{j}u,
\end{equation*}%
where $\nabla _{i}$ is the \emph{covariant derivative}
(Levi--Civita connection, see Appendix). We say that a $C^{2}$
function $u:M\times \lbrack 0,T)\rightarrow \mathbb{R},$ where
$T\in (0,\infty ],$ is a solution to the heat equation if
(\ref{h1}) holds. One of the most important properties satisfied
by the heat equation is the \textit{maximum principle}, which says
that for any smooth solution to the heat equation, whatever
point-wise bounds hold at $t=0$ also hold for $t>0$
\cite{CaoChow}. More precisely, we can state: Let $u:M\times
\lbrack 0,T)\rightarrow \mathbb{R}$ be a $C^{2}$ solution to the
heat equation (\ref{h1}) on a complete Riemannian $3-$manifold
$M$. If $C_{1}\leq u\left( x,0\right) \leq C_{2}$ for all $x\in
M,$ for some constants $C_{1},C_{2}\in \mathbb{R},$ then
$C_{1}\leq u\left( x,t\right) \leq C_{2}$ for all $x\in M$ and
$t\in \lbrack 0,T).$ This property exhibits the averaging behavior
of the heat equation (\ref{h1}) on $M$.

Now, consider Perelman's \emph{entropy functional} \cite{Perel1}
on a 3--manifold $M\footnote{%
Note that in the related context of Riemannian gravitation theory, the
so--called \emph{gravitational entropy} is embedded in the Weyl curvature $%
(4,0)-$tensor $\mathfrak{W}$, which is the traceless component of the
Riemann curvature tensor $\mathfrak{Rm}$ (i.e., $\mathfrak{Rm}$ with the
Ricci tensor $\mathfrak{Rc}$ removed),
\begin{equation*}
\mathfrak{W}=\mathfrak{Rm}-f(R_{ij}g_{ij}),
\end{equation*}%
where $f(R_{ij}g_{ij})$ is a certain linear function of $R_{ij}$ and $g_{ij}$%
. According to Penrose's \emph{Weyl curvature hypothesis}, the entire \emph{%
history of a closed universe} starts from a uniform low--entropy Big Bang
with zero Weyl curvature tensor of the cosmological gravitational field and
ends with a high--entropy Big Crunch, representing the congealing of may
black holes, with Weyl tensor approaching infinity (see \cite%
{Penr79,HawkPenr}).}$
\begin{equation}
\mathcal{F}=\int_{M}(R+|\nabla f|^{2}){\mathrm{e}}^{-f}d\mu  \label{F}
\end{equation}%
for a Riemannian metric $g_{ij}$ and a (temperature-like) scalar function $f$ (which satisfies the backward heat equation)
on a closed 3--manifold $M$, where $d\mu $ is the volume 3--form (\ref{dmu}%
). During the Ricci flow (\ref{RF}), $\mathcal{F}$ evolves on $M$ as
\begin{equation}
\partial _{t}\mathcal{F}=2\int |R_{ij}+\nabla _{i}\nabla _{j}f|^{2}{\mathrm{e%
}^{-f}}d\mu {.}  \label{dF}
\end{equation}%
Now, define $\lambda (g_{ij})=\inf \mathcal{F}(g_{ij},f),$ where infimum is
taken over all smooth $f,$ satisfying
\begin{equation}
\int_{M}{\mathrm{e}^{-f}}d\mu =1.  \label{eDm}
\end{equation}%
$\lambda (g_{ij})$ is the lowest eigenvalue of the operator $-4\Delta +R.$
Then the entropy evolution formula (\ref{dF}) implies that $\lambda
(g_{ij}(t))$ is nondecreasing in $t,$ and moreover, if $\lambda
(t_{1})=\lambda (t_{2}),$ then for $t\in \lbrack t_{1},t_{2}]$ we have $%
R_{ij}+\nabla _{i}\nabla _{j}f=0$ for $f$ which minimizes $\mathcal{F}$ on $%
M $ \cite{Perel1}. Thus a steady breather on $M$ is necessarily a steady
soliton.

If we define the conjugate \emph{heat operator} on $M$ as
\begin{equation*}
\Box ^{\ast }=-\partial /\partial t-\Delta +R
\end{equation*}%
then we have the \emph{conjugate heat equation}\footnote{%
In \cite{Perel1} Perelman stated a differential Li--Yau--Hamilton (LYH) type
inequality \cite{Hsu2} for the fundamental solution $u=u(x,t)$ of the
conjugate heat equation (\ref{conHeat1}) on a closed $n-$manifold $M$
evolving by the Ricci flow (\ref{RF}). Let $p\in M$ and
\begin{equation*}
u=(4\pi \tau )^{-\frac{n}{2}}\mathrm{e}^{-f}
\end{equation*}%
be the fundamental solution of the conjugate heat equation in $M\times (0,T)$%
,
\begin{equation*}
\Box ^{\ast }u=0,\qquad\text{or}\qquad\partial _{t}u+\Delta u=Ru,
\end{equation*}%
where $\tau =T-t$ and $R=R(\cdot ,t)$ is the scalar curvature of $M$ with
respect to the metric $g(t)$ with $\lim_{t\nearrow T}u=\delta _{p}$ (in the
distribution sense), where $\delta _{p}$ is the delta--mass at $p$. Let
\begin{equation*}
v=[\tau (2\Delta f-|\nabla f|^{2}+R)+f-n]u,
\end{equation*}%
where $\tau =T-t$. Then we have a differential LYH--type inequality
\begin{equation}
v(x,t)\leq 0\quad \text{ in \ }M\times (0,T).  \label{ineq1}
\end{equation}%
This result was used by Perelman to give a proof of the \textit{%
pseudolocality theorem} \cite{Perel1} which roughly said that almost
Euclidean regions of large curvature in closed manifold with metric evolving
by Ricci flow $g(t)$ given by (\ref{RF}) remain localized.
\par
In particular, let $(M,g(t))$, $0\leq t\leq T$, $\partial M\neq \phi $, be a
compact $3-$manifold with metric $g(t)$ evolving by the Ricci flow $g(t)$
given by (\ref{RF}) such that the second fundamental form of the surface $%
\partial M$ with respect to the unit outward normal $\partial /\partial \nu $
of $\partial M$ is uniformly bounded below on $\partial M\times \lbrack 0,T]$%
. A global Li--Yau gradient estimate \cite{LY}\ for the solution of the
generalized conjugate heat equation was proved in \cite{Hsu2} (using a a
variation of the method of P.~Li and S.T.~Yau, \cite{LY}) on such a manifold
with Neumann boundary condition.} \cite{Perel1}
\begin{equation}
\Box ^{\ast }u=0.  \label{conHeat1}
\end{equation}

The entropy functional (\ref{F}) is nondecreasing under the coupled \emph{%
Ricci--diffusion flow} on $M$ (see \cite{Y2,Li07})
\begin{equation}
\partial _{t}g_{ij}=-2R_{ij},\qquad \partial _{t}u=-\Delta u+\frac{R}{2}u-%
\frac{|\nabla u|^{2}}{u},  \label{conHeat}
\end{equation}%
where the second equation ensures ~ $ \int_{M}u^{2}d\mu =1, $~
to be preserved by the Ricci flow $g(t)$ on $M$. If we define $\ u=\mathrm{e}%
^{-\frac{f}{2}}$, then the right--hand equation in (\ref{conHeat})
is equivalent to the generic scalar--field $f-$evolution equation
on $M$,
\begin{equation*}
\partial _{t}f=-\Delta f-R+|\nabla f|^{2},
\end{equation*}%
which instead preserves (\ref{eDm}).

The coupled Ricci--diffusion flow (\ref{conHeat}), or
equivalently, the dual system
\begin{equation}
\partial _{t}g_{ij}=\Delta _{M}g_{ij}+Q_{ij}(g,\partial g), \qquad \partial _{t}f=-\Delta f-R+|\nabla
f|^{2},\label{dual}
\end{equation}
is our \emph{global decay} model for a general $n-$dimensional
bio--reaction--diffusion process, including both geometric and
bio--chemical multi--phase evolution.

The sole \emph{hypothesis of this paper} is that any kind of
reaction--diffusion processes in biology, chemistry and physics is
subsumed by the geometric--diffusion system (\ref{conHeat}), or
the dual system (\ref{dual}).

\section{Conclusion}

In this paper we have conjectured that the Ricci-flow equation from Riemannian geometry: $$\partial _{t}g_{ij}=\Delta _{M}g_{ij}+Q_{ij}(g,\partial g),$$
can be considered as a general geometric framework for various nonlinear
reaction-diffusion systems (and related dissipative solitons) in mathematical biology.
More precisely, we proposed a hypothesis that any kind of reaction-diffusion processes in biology, chemistry and physics can be modelled by the combined geometric-diffusion system represented by the Ricci-flow equation. The validity of this hypothesis was demonstrated by reviewing a number of popular nonlinear reaction-diffusion systems from biology, chemistry and physics and showed that they could all be subsumed by the geometric framework of the Ricci flow.

\section{Appendix: Riemann and Ricci curvatures on a smooth $n-$manifold}

Recall that proper differentiation of vector and tensor fields on a smooth
Riemannian $n-$manifold is performed using the \textit{Levi--Civita
covariant derivative} (see, e.g., \cite{GaneshSprBig,GaneshADG}). Formally,
let $M$ be a Riemannian $n-$manifold with the tangent bundle $TM$ and a
local coordinate system $\{x^{i}\}_{i=1}^{n}$ defined in an open set $%
U\subset M$. The covariant derivative operator, $\nabla _{X}:C^{\infty
}(TM)\rightarrow C^{\infty }(TM)$, is the unique linear map such that for
any vector fields $X,Y,Z,$ constant $c$, and function $f$ the following
properties are valid:
\begin{eqnarray*}
\nabla _{X+cY} &=&\nabla _{X}+c\nabla _{Y}, \\
\nabla _{X}(Y+fZ) &=&\nabla _{X}Y+(Xf)Z+f\nabla _{X}Z, \qquad \text{with}\\
\nabla _{X}Y-\nabla _{Y}X &=&[X,Y],\qquad \text{(torsion free property)}
\end{eqnarray*}%
where $[X,Y]$ is the Lie bracket of $X$ and $Y$ (see, e.g., \cite{VladSiam}%
). In local coordinates, the metric $g$ is defined for any orthonormal basis
$(\partial_i=\partial_{x^i})$ in $U\subset M$ by
\begin{equation*}
g_{ij}=g(\partial_{i},\partial_{j})=\delta _{ij}, \qquad
\partial_{k}g_{ij}=0.
\end{equation*}
Then the affine \textit{Levi--Civita connection} is defined on $M$
by
\begin{equation*}
\nabla _{\partial _{i}}\partial _{j}=\Gamma _{ij}^{k}\partial
_{k},\qquad \text{where\qquad }\Gamma
_{ij}^{k}=\frac{1}{2}g^{kl}\left( \partial _{i}g_{jl}+\partial
_{j}g_{il}-\partial _{l}g_{ij}\right)
\end{equation*}%
are the (second-order) \textit{Christoffel symbols}.

Now, using the covariant derivative operator $\nabla _{X}$ we can define the
\textit{Riemann curvature} $(3,1)-$tensor $\mathfrak{Rm}$ by (see, e.g.,
\cite{GaneshSprBig,GaneshADG})
\begin{equation*}
\mathfrak{Rm}(X,Y)Z=\nabla _{X}\nabla _{Y}Z-\nabla _{Y}\nabla _{X}Z-\nabla
_{\lbrack X,Y]}Z.
\end{equation*}%
$\mathfrak{Rm}$ measures the curvature of the manifold by expressing how
noncommutative covariant differentiation is. The $(3,1)-$components $%
R_{ijk}^{l}$ of $\mathfrak{Rm}$ are defined in $U\subset M$ by
\begin{eqnarray*}
\mathfrak{Rm}\left( \partial _{i},\partial _{j}\right) \partial _{k}
&=&R_{ijk}^{l}\partial _{l},\qquad\text{which expands (see \cite{MTW}) as} \\
R_{ijk}^{l} &=&\partial _{i}\Gamma _{jk}^{l}-\partial _{j}\Gamma
_{ik}^{l}+\Gamma _{jk}^{m}\Gamma _{im}^{l}-\Gamma _{ik}^{m}\Gamma
_{jm}^{l}.
\end{eqnarray*}%
Also, the Riemann $(4,0)-$tensor $R_{ijkl}=g_{lm}R_{ijk}^{m}$ is defined as
the $g-$based inner product on $M$,
\begin{equation*}
R_{ijkl}=\left\langle \mathfrak{Rm}\left( \partial _{i},\partial _{j}\right)
\partial _{k},\partial _{l}\right\rangle .
\end{equation*}

The first and second Bianchi identities for the Riemann $(4,0)-$tensor $%
R_{ijkl}$ hold,
\begin{equation*}
R_{ijkl}+R_{jkil}+R_{kijl}=0,\qquad \nabla _{i}R_{jklm}+\nabla
_{j}R_{kilm}+\nabla _{k}R_{ijlm}=0,
\end{equation*}
while the twice contracted second Bianchi identity reads
\begin{equation}
2\nabla _{j}R_{ij}=\nabla _{i}R.  \label{cB}
\end{equation}

The $(0,2)$ \textit{Ricci tensor} $\mathfrak{Rc}$ is the trace of
the Riemann $(3,1)-$tensor $\mathfrak{Rm}$,
\begin{equation*}
\mathfrak{Rc}(Y,Z)+\mathrm{tr}(X\rightarrow
\mathfrak{Rm}(X,Y)Z),\qquad \text{so that\qquad
}\mathfrak{Rc}(X,Y)=g(\mathfrak{Rm}(\partial _{i},X)\partial
_{i},Y),
\end{equation*}
Its components $R_{jk}=\mathfrak{Rc}\left( \partial _{j},\partial
_{k}\right) $ are given in $U\subset M$ by the contraction
\cite{MTW}
\begin{eqnarray*}
R_{jk} &=&R_{ijk}^{i},\qquad \text{or, ~in terms of Christoffel symbols,} \\
R_{jk} &=&\partial _{i}\Gamma _{jk}^{i}-\partial _{k}\Gamma _{ji}^{i}+\Gamma
_{mi}^{i}\Gamma _{jk}^{m}-\Gamma _{mk}^{i}\Gamma _{ji}^{m}.
\end{eqnarray*}
Being a symmetric second--order tensor, $\mathfrak{Rc}$ has ${n+1}{2}$ independent components on an $n-$manifold $M$. In particular, on a
3--manifold, it has 6 components, and on a 2--surface it has only the
following 3 components:
\begin{equation*}
R_{11}=g^{22}R_{2112},\qquad R_{12}=g^{12}R_{2121},\qquad
R_{22}=g^{11}R_{1221},
\end{equation*}
which are all proportional to the corresponding coordinates of the metric
tensor,
\begin{equation}
\frac{R_{11}}{g_{11}}=\frac{R_{12}}{g_{12}}=\frac{R_{22}}{g_{22}}=-\frac{%
R_{1212}}{\det (g)}.  \label{Rsrf}
\end{equation}

Finally, the scalar curvature $R$ is the trace of the Ricci tensor $%
\mathfrak{Rc}$, given in $U\subset M$ by:~ $ R=g^{ij}R_{ij}. $


\begin{thebibliography}{99}
\bibitem{Ablowitz} \textsc{M.J. Ablowitz, D.J. Kaup, A.C. Newell, and H.
Segur}, Method for solving the sine-Gordon equation. Phys. Rev. Let. {30}%
(1973), pp. 1262--1264.

\bibitem{Akhmediev} \textsc{N.N. Akhmediev, V.M. Eleonskii, and N.E. Kulagin}%
, First-order exact solutions of the nonlinear Schr\"odinger equation. Th.
Math. Physics {72}(1987), pp. 809--818.

\bibitem{Amari} \textsc{S. Amari}, Dynamics of pattern formation in
lateral-inhibition type neural fields. Biol. Cybern. {27}(1977), pp. 77--87.

\bibitem{Anderson} \textsc{M.T. Anderson}, Geometrization of 3-manifolds via
the Ricci flow, Not. Am. Math. Soc. {51}2(2004), pp. 184-193.

\bibitem{Barkley} \textsc{D. Barkley}, A model for fast computer--simulation
of waves in excitable media. Physica D {49}(1991), pp. 61--70.

\bibitem{Bode} \textsc{M. Bode}, Front-bifurcations in reaction-diffusion
systems with inhomogeneous parameter distributions, Physica D {106}(1997),
pp. 270--286.

\bibitem{CaoChow} \textsc{H.D. Cao and B. Chow}, Recent developments on the
Ricci flow, Bull. Amer. Math. Soc. {36}(1999), pp. 59-74.

\bibitem{ChowKnopf} \textsc{B. Chow and D. Knopf}, The Ricci flow: An
introduction, Mathematical Surveys and Monographs, AMS, Providence, RI,
(2004).

\bibitem{Conway08} \textsc{J.M. Conway and H. Riecke}, Superlattice Patterns
in the Complex Ginzburg-Landau Equation with Multi-Resonant
Forcing, arXiv:nlin.PS.0803.0346, (2008).

\bibitem{Field07} \textsc{R.J. Field}, Oregonator,
Scholarpedia, {2}:5(2007), pp. 1386.

\bibitem{FKNorig} \textsc{R.J. Field, E. K"{o}r"{o}s, and R.M. Noyes},
Oscillations in chemical systems. J. Amer. Chem. Soc. textbf{94}(1972), pp.
8649--8664.

\bibitem{Fitzhugh} \textsc{R. FitzHugh}, Impulses and physiological states
in theoretical models of nerve membrane. Biophys. J. {1}(1961), pp. 445-466.

\bibitem{Friedrich} \textsc{R. Friedrich}, Group Theoretic Methods in the
Theory of Pattern Formation, in Collective dynamics of nonlinear and
disordered systems, Springer, Berlin, (2004).

\bibitem{GM72} \textsc{A. Gierer and H. Meinhardt}, A theory of biological
pattern formation. Kybern. {12}(1972), pp. 30-39.

\bibitem{Gurevich} \textsc{S.V. Gurevich, S. Amiranashvili, and H.-G. Purwins%
}, Breathing dissipative solitons in three-component reaction-diffusion
system. Phys. Rev. E{74}(2006), pp. 066201.

\bibitem{Ham82} \textsc{R.S. Hamilton}, Three-manifolds with positive Ricci
curvature, J. Diff. Geom. {17}(1982), pp. 255-306.

\bibitem{4-manifold} \textsc{R.S. Hamilton}, Four-manifolds with positive
curvature operator, J. Dif. Geom. {24}(1986), pp. 153-179.

\bibitem{surface} \textsc{R.S. Hamilton}, The Ricci flow on surfaces, Cont.
Math. {71}(1988), pp. 237-261.

\bibitem{Harnack} \textsc{R.S. Hamilton}, The Harnack estimate for the Ricci
flow, J. Dif. Geom. {37}(1993), pp. 225-243.

\bibitem{non-singular} \textsc{R.S. Hamilton}, Non-singular solutions of the
Ricci flow on three-manifolds, Comm. Anal. Geom., {7}:4(1999), pp. 695-729.

\bibitem{HawkPenr} \textsc{S. Hawking and R. Penrose}, The Nature of Space
and Time, Princeton Univ. Press, (1996).

\bibitem{Hsu2} \textsc{S.Y. Hsu}, Some results for the Perelman LYH-type
inequality, arXiv:math.DG/0801.3506, (2008).

\bibitem{VladSiam} \textsc{V. Ivancevic}, Symplectic Rotational Geometry in
Human Biomechanics, SIAM Rev. {46}:3(2004), pp. 455--474.

\bibitem{GaneshWSci} \textsc{Ivancevic, V. and Ivancevic, T.}, Natural
Biodynamics. World Scientific, Singapore, (2006).

\bibitem{GaneshSprBig} \textsc{V. Ivancevic and T. Ivancevic}, Geometrical
Dynamics of Complex Systems. Springer, Dordrecht, (2006).

\bibitem{StrAttr} \textsc{V. Ivancevic and T. Ivancevic}, High--Dimensional
Chaotic and Attractor Systems. Springer, Berlin, (2006).

\bibitem{GCompl} \textsc{V. Ivancevic and T. Ivancevic}, Complex Dynamics:
Advanced System Dynamics in Complex Variables. Springer, Dordrecht, (2007).

\bibitem{GaneshADG} \textsc{V. Ivancevic and T. Ivancevic}, Applied
Differfential Geometry: A Modern Introduction. World Scientific, Singapore,
(2007).

\bibitem{NeuFuz} \textsc{V. Ivancevic and T. Ivancevic}, Neuro-Fuzzy
Associative Machinery for Comprehensive Brain and Cognition Modelling.
Springer, Berlin, (2007).

\bibitem{CompMind} \textsc{V. Ivancevic and T. Ivancevic}, Computational
Mind: A Complex Dynamics Perspective. Springer, Berlin, (2007).

\bibitem{Complexity} \textsc{V. Ivancevic and T. Ivancevic}, Complex
Nonlinearity: Chaos, Phase Transitions, Topology Change and Path Integrals.
Springer, Berlin, (2008).

\bibitem{QuLeap} \textsc{V. Ivancevic and T. Ivancevic}, Quantum Leap: From
Dirac and Feynman, Across the Universe, to Human Body and Mind. World
Scientific, Singapore, (2008).

\bibitem{TacaNODY} \textsc{T. Ivancevic, L. Jain, J. Pattison, and A. Hariz}%
, Nonlinear Dynamics and Chaos Methods in Neurodynamics and
Complex Data Analysis. Nonl. Dyn. (to appear, On line first,
Springer).

\bibitem{kve2} \textsc{A.~Kaminaga, V.K.~Vanag, and I.R.~Epstein}, ``Black
spots" in a surfactant-rich Belousov-Zhabotinsky reaction dispersed in a
water-in-oil microemulsion system. J. Chem. Phys. {122}(2005), pp. 174706.

\bibitem{kve} \textsc{A.~Kaminaga, V.K.~Vanag, and I.R.~Epstein}, A
reaction-diffusion memory device. Ang. Chem. {45}(2006), pp. 3087.

\bibitem{Kohler} \emph{G. Kohler and C. Milstein}, Continuous cultures of
fused cells secreting antibody of predefined specificity. Nature, 256(1975),
pp. 495.

\bibitem{Kolokolnikov} \textsc{T. Kolokolnikov and M. Tlidi}, Spot
deformation and replication in the two-dimensional Belousov-Zhabotinsky
reaction in water-in-oil microemulsion, Phys. Rev. Lett. {98}(2007), pp.
188303.

\bibitem{Kunz} \textsc{L.A. Kunz-Schughart}, Multicellular tumor spheroids:
intermediates between monolayer culture and in vivo tumor, Cell Biol. Int. {%
23}:3(1999), pp. 157-61.

\bibitem{LY} \textsc{P. Li and S.T. Yau}, On the parabolic kernel of the Schr%
\"{o}dinger operator. Acta Math. {156}(1986), pp. 153--201.

\bibitem{Li07} \textsc{J. Li}, First variation of the Log Entropy functional
along the Ricci flow, arXiv:math.DG/0712.0832, (2007).

\bibitem{Mackenzie} \textsc{D. Mackenzie}, Perelman Declines Math's Top
Prize; Three Others Honored in Madrid, Science 313, pp. 1027, (2006).

\bibitem{GM06} \textsc{H. Meinhardt}, Gierer-Meinhardt model, Scholarpedia, {%
1}:12(2006), pp. 1418.

\bibitem{Milnor} \textsc{J. Milnor}, Towards the Poincar\'{e} Conjecture and
the Classification of 3-Manifolds, Not. Am. Math. Soc. {50}:10(2003), pp.
1226-1233.

\bibitem{MTW} \textsc{C. Misner, K. Thorne, and J.A. Wheeler}, Gravitation,
W.H. Freeman and Company, (1973).

\bibitem{Missel} \textsc{A.R. Missel and K.A. Dahmen}, Hopping Conduction
and Bacteria: Transport in Disordered Reaction-Diffusion Systems, Phys. Rev.
Let. {100}(2007), pp. 058301.

\bibitem{Morgan} \textsc{S.W. Morgan, I.V. Biktasheva, and V.N. Biktashev},
Control of scroll wave turbulence using resonant perturbations.
arXiv:nlinPS.0806.2262, (2008).

\bibitem{Nagumo} \textsc{J. Nagumo, S. Arimoto, S. Yoshizawa}, An active
pulse transmission line simulating nerve axon. Proc IRE. 50(1962), pp.
2061--2070.

\bibitem{Nelson98} \textsc{D.R. Nelson and N.M. Shnerb}, Non-hermitian
localization and population biology. Phys. Rev. E {58}(1998), pp. 1383--1403.

\bibitem{HopfQuench} \textsc{K. Nielsen, F. Hynne, P.G. Sorensen}, Hopf
bifurcation in chemical kinetics, J. Chem. Phys. textbf{94}(1991), pp.
1020--1029.

\bibitem{Penr79} \textsc{R. Penrose}, Singularities and Time-Asymmetry, in
General Relativity: An Einstein Centenary Survey, (ed. S. Hawking, W.
Israel), 581-638, Cambridge Univ. Press, (1979).

\bibitem{Perel1} \textsc{G. Perelman}, The entropy formula for the Ricci
flow and its geometric applications, arXiv:math.DG/0211159, (2002).

\bibitem{Perel2} \textsc{G. Perelman}, Ricci flow with surgery on
three-manifolds, arXiv:math.DG/0303109, (2003).

\bibitem{Preziosi} \textsc{L. Preziosi}, Cancer Modeling and Simulation. CRC
Press, (2003).

\bibitem{Pri80} \textsc{I. Prigogine}, From Being to Becoming: Time and
Complexity in the Physical Sciences. Freeman, San Francisco, (1980).

\bibitem{Purwins} \textsc{H.-G. Purwins, H.U. Bodeker, and A.W. Liehr},
Dissipative Solitons in Reaction-Diffusion Systems, in Dissipative Solitons
(ed. N. Akhmediev, A. Ankiewicz), Lecture Notes in Physics, Springer, (2005).

\bibitem{Rabi} \textsc{M.I. Rabinovich, A.B. Ezersky, and P.D. Weidman}, The
Dynamics of Patterns, World Scientific, Singapore, (2000).

\bibitem{Roose} T. Roose, S.J. Chapman, P.K. Maini, Mathematical Models of
Avascular Tumor Growth. SIAM Rev. 49:2(2007), pp. 179--208.

\bibitem{Schoner} \textsc{G. Sch\"{o}ner}, Dynamical Systems Approaches to
Cognition. In: Cambridge Handbook of Computational Cognitive Modeling.
Cambridge University Press. R. Sun (ed), (2007).

\bibitem{Sutherland} \textsc{R.M. Sutherland}, Cell and environment
interactions in tumor microregions: The multicell spheroid model, Science, {%
240}(1988), pp. 177--184.

\bibitem{Thurston} \textsc{W. Thurston}, Three-dimensional manifolds,
Kleinian groups and hyperbolic geometry, Bull. Amer. Math. Soc. {6}(1982),
pp. 357-381.

\bibitem{Turing} \textsc{A.M. Turing}, The Chemical Basis of Morphogenesis.
Phil. Trans. Roy. Soc. London, B {237}(1952), pp. 37--72.

\bibitem{Tyson} \textsc{J.J. Tyson}, A quantitative account of oscillations,
bistability and travelling waves in the Belousov-Zhabotinskii reaction, in
Oscillation and Travelling Waves in Chemical Systems, eds. Field, R.~J.,
Burger, M., Wiley-Intersc., New York, (1985).

\bibitem{Yau} \textsc{S.T. Yau}, Structure of Three-Manifolds --- Poincar%
\'{e} and geometrization conjectures. talk given at the Morningside Center
of Mathematics on June 20, (2006).

\bibitem{Y2} \textsc{R. Ye}, The log entropy functional along the Ricci
flow. arXiv:math.DG/0708.2008v3, (2007).

\bibitem{Zhabotinsky} \textsc{A.M. Zhabotinsky}, Belousov--Zhabotinsky
reaction, Scholarpedia, {2}:9(2007), pp. 1435.
\end{thebibliography}
\end{document}